\documentclass[preprint]{aastex}
\usepackage{natbib}
\usepackage{graphicx}
\newcommand{\hst}{\textit{HST}}
\newcommand{\lam}{$\lambda$}
\newcommand{\ib}{\,{\sc i}}
\newcommand{\ii}{\,{\sc ii}}
\newcommand{\iii}{\,{\sc iii}}
\newcommand{\iv}{\,{\sc iv}}
\newcommand{\vb}{\,{\sc v}}

\shorttitle{Super Star Cluster NGC 3125-1}
\shortauthors{Wofford et al.}
\begin{document}
\title{A Rare Encounter with Very Massive Stars in NGC 3125-A1}
\author{Aida Wofford\altaffilmark{1}, Claus Leitherer\altaffilmark{2}, Rupali Chandar\altaffilmark{3}, and Jean-Claude Bouret\altaffilmark{4}}
\affil{\altaffilmark{1}UPMC-CNRS, UMR7095, Institut d'Astrophysique de Paris, F-75014, Paris, France}
\email{wofford@iap.edu}
\affil{\altaffilmark{2}Space Telescope Science Institute, 3700 San Martin Drive, Baltimore, MD 21218}
\affil{\altaffilmark{3}University of Toledo, Department of Physics and Astronomy, Toledo, OH 43606}
\affil{\altaffilmark{4}Aix Marseille Universite, CNRS, LAM (Laboratoire d'Astrophysique de Marseille) UMR 7326, 13388, Marseille, France}

\begin{abstract}
Super star cluster A1 in the nearby starburst galaxy NGC 3125 is characterized by broad  He\ii~\lam1640 emission (full width at half maximum, $FWHM\sim1200$ km s$^{-1}$) of unprecedented strength (equivalent width, $EW=7.1\pm0.4$ \AA). Previous attempts to characterize the massive star content in NGC 3125-A1 were hampered by the low resolution of the UV spectrum and the lack of co-spatial panchromatic data. We obtained far-UV to near-IR spectroscopy of the two principal emitting regions in the galaxy with the Space Telescope Imaging Spectrograph (STIS) and the Cosmic Origins Spectrograph (COS) onboard the Hubble Space Telescope (\hst). We use these data to study three clusters in the galaxy, A1, B1, and B2. We derive cluster ages of 3-4 Myr, intrinsic reddenings of $E(B-V)=0.13$, 0.15, and 0.13, and cluster masses of $1.7\times10^5$, $1.4\times10^5$, and $1.1\times10^5$ M$_\odot$, respectively. A1 and B2 show O\vb~\lam1371 absorption from massive stars, which is rarely seen in star-forming galaxies, and have Wolf-Rayet (WR) to O star ratios of $N(WN5-6)/N(O)=0.23$ and 0.10, respectively. The high $N(WN5-6)/N(O)$ ratio of A1 cannot be reproduced by models that use a normal IMF and generic WR star line luminosities. We rule out that the extraordinary He\ii~\lam1640 emission and O\vb~\lam1371 absorption of A1 are due to an extremely flat upper IMF exponent, and suggest that they originate in the winds of very massive ($>120\,M_\odot$) stars. In order to reproduce the properties of peculiar clusters such as A1, the present grid of stellar evolution tracks implemented in Starburst99 needs to be extended to masses $>120\,M_\odot$. 
\end{abstract}

\section{INTRODUCTION}

Starburst galaxies are a class of objects which display characteristics associated with massive violent bursts of star formation. Locally galaxies with active star formation are responsible for ~20\% of the entire high-mass star formation \citep{bri04}. Nearby starburst galaxies are the most obvious local counterparts of the normal star-forming galaxies discovered at high redshift (e.g., \citealt{ste96, ste99}). Using large spectroscopic datasets of these $z\sim3$ Lyman break galaxies (LBGs), high S/N composite spectra have been obtained, which provide a glimpse of the dominant stellar populations present in the early universe (e.g, \citealt{ste03, sha03}).

Starburst galaxies are powered by populations of hot stars with progenitor masses of~5\,M$_\odot$ and above. Such stars have almost all their diagnostic lines in the satellite UV so that space observations are necessary for understanding their nature at low redshift. Figure 7 of \cite{lei11} shows the UV composite spectrum of 46 nearby starburst galaxies. The most prominent features are the N\vb~1240, Si\iv~1400, and C\iv~1550 resonance doublets \citep{lei01}. These UV lines originate in powerful stellar winds with stellar-mass-dependent properties and cover a wide range of ionization potentials from 13.6 eV (O\,{\sc i}) to 87 eV (O\vb). This makes them suitable for studying the mass distribution and eventually the initial mass function (IMF) of the most massive stars in the mass range between 10 and 100\,M$_\odot$ (upper IMF). A major outcome of numerous spectroscopic UV studies of local star-forming galaxies is the ubiquity of a single Salpeter-like IMF in this mass range \citep{lei09,bas10}.

Observations of some local starbursts also show broad He\ii~$\lambda$1640 emission (\citealt{cha04}, hereafter C04). The fact that the line is broad ($FWHM~\sim10^3$ km s$^{-1}$) excludes a nebular origin but suggests stellar winds as the formation mechanism. In order to produce detectable stellar He\ii~emission, one or more of the following conditions must be met: the stars must be hot to doubly ionize He, the mass-loss rates must be high to produce dense winds, and He must be overabundant. Since WR stars meet all three conditions, the presence of He\ii~\lam1640 is generally thought to indicate WR stars \citep{con83, hum85, cro07}. He\ii~\lam1640 is also present in rare but luminous OIf supergiants. However, in these stars, it is significantly narrower and weaker than in the early WN and WC Wolf-Rayet subtypes. On the other hand, OIf stars have similar He\ii~\lam1640 widths to late WN and WC stars, albeit weaker emission \citep{wal95,bou12,cro98,cro11}. Additionally, extremely massive H-burning O stars can develop spectra similar to those of WR stars \citep{cro11}. Therefore He\ii~\lam1640 may be related to a very young ($< $3 Myr) star cluster containing an excess of extremely massive stars close to the main-sequence. He\ii~\lam1640 is normally weak in local starbursts, as expected for a normal IMF and given the short life-times of WR stars. On the other hand, broad (non-nebular) He\ii~$\lambda$1640 is the strongest stellar line in the composite spectrum of 811 LBGs created by \cite{sha03}. This is the only strongly  discrepant stellar line when the rest-frame UV spectra of LBGs are compared to the UV templates of local starbursts, in particular those computed with the widely used stellar population synthesis package Starburst99 \citep{lei99, lei10}.

The nearby WR galaxy NGC 3125 can be used as a template to interpret the data of higher redshift starbursts (C04). This galaxy hosts a compact star cluster, NGC 3125-A1 that is characterized by exceptionally broad ($FWHM=1400$ km s$^{-1}$) and strong ($EW=7.1$ \AA) He\ii~$\lambda$1640 emission (C04, they call the cluster NGC 3125-1). Such an equivalent width is by far the largest value observed at any redshift, as it is $\sim3$ times the mean of local starburst galaxies ($EW=2.7$ \AA, C04) and three times stronger than the largest $EW$ observed in galaxies located at $z>2$ ($EW=2.7$~\AA, \citealt{erb10}). For the composite spectrum of LBGs analyzed by \cite{sha03}, $EW=1.3\pm0.3$ \AA~as measured by \cite{bri08}. Another peculiarity of NGC 3125-A1 is its strong O\vb~\lam1371 absorption. This feature is absent in the composite starburst spectrum of \cite{lei11} and is observed in the most massive stars in the local universe \citep{wal85}. Finally, C04 found an extreme WR-to-O star ratio of $>1$ in cluster A1, while the predicted value by standard stellar evolution models is $\sim0.1$ for an instantaneous burst \citep{sch99}. If such clusters were present in large numbers in LBGs, the LBG rest-frame UV spectra could be empirically reproduced and understood. This leaves open the question why such unique clusters are rare in the local universe and why they could be abundant in the early universe. Complicating the interpretation of NGC 3125-A1 is the lack of co-spatial panchromatic data. C04 had to rely on large-aperture ground-based data for complementary stellar and nebular lines in their spectral synthesis. The resulting uncertainties, in particular the reddening correction, made the interpretation of an excess of WR stars rather uncertain, as correctly pointed out by \cite{had06}, hereafter H06. From a combined UV and optical study, the latter authors derived a more modest WR-to-O ratio of 0.2.

Given the potentially unique nature of NGC 3125, we obtained co-spatial FUV-to-NIR spectroscopy of the two main emission regions in this galaxy, and UV spectroscopy at higher spectral resolution for region A1. Our principal goal is to once and for all settle the remaining questions on cluster A1 and to establish or reject its extraordinary nature. These data have implications for understanding the rest-frame UV spectra of high redshift ($z>2$) galaxies. They are also useful for calibrating model spectral energy distributions (SEDs), which are the basis of most mass and star formation rate determinations at high redshifts. In section 2 we present the target and describe the data and data reduction. In section 3 we analyze the data. In section 4 we discuss our results. Finally, in section 5 we provide a summary and conclusion.

\section{TARGET, OBSERVATIONS, \& DATA REDUCTION}

\subsection{NGC 3125}

\begin{figure}
\rotate
\epsscale{0.48}
\plotone{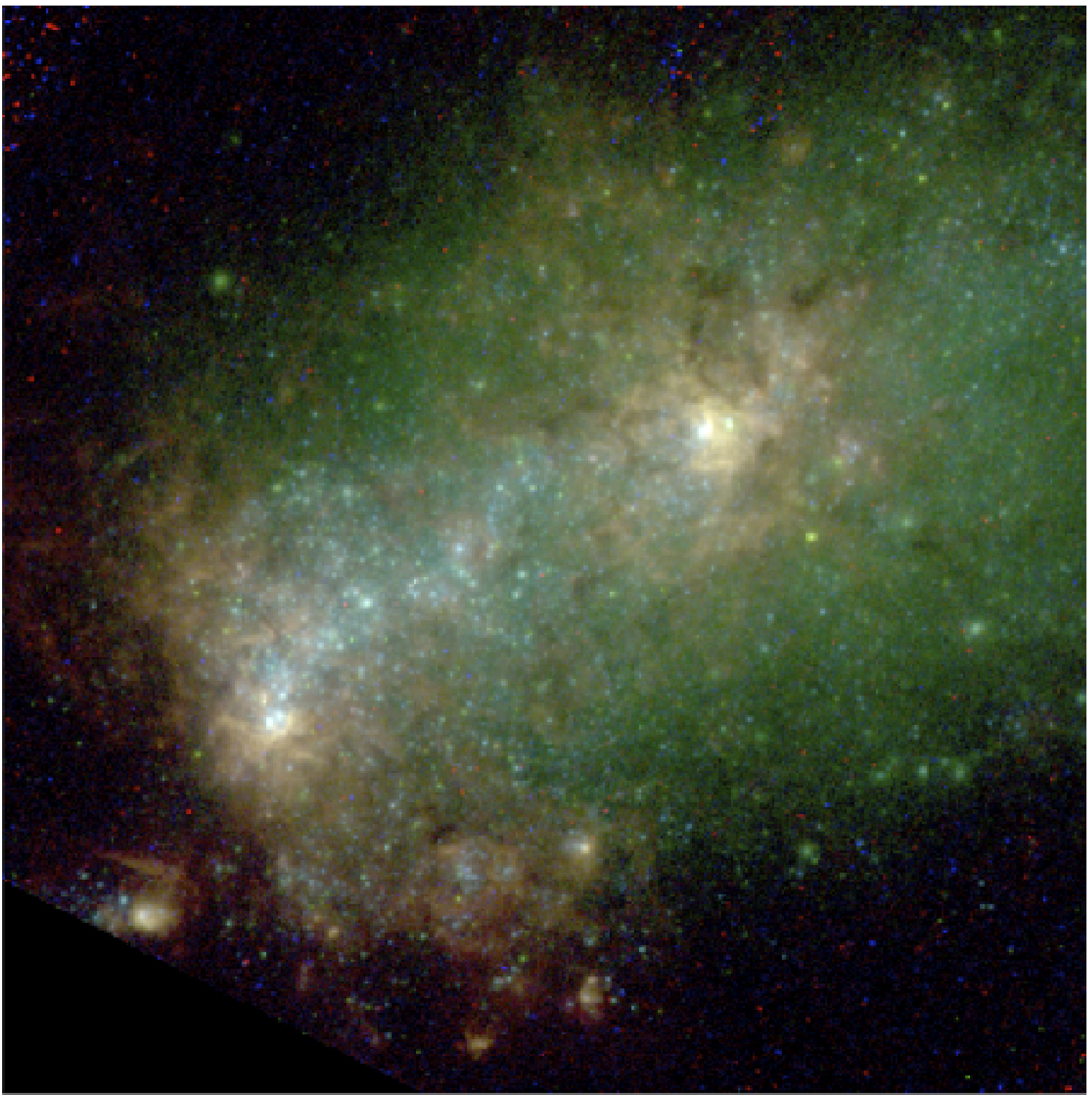}\plotone{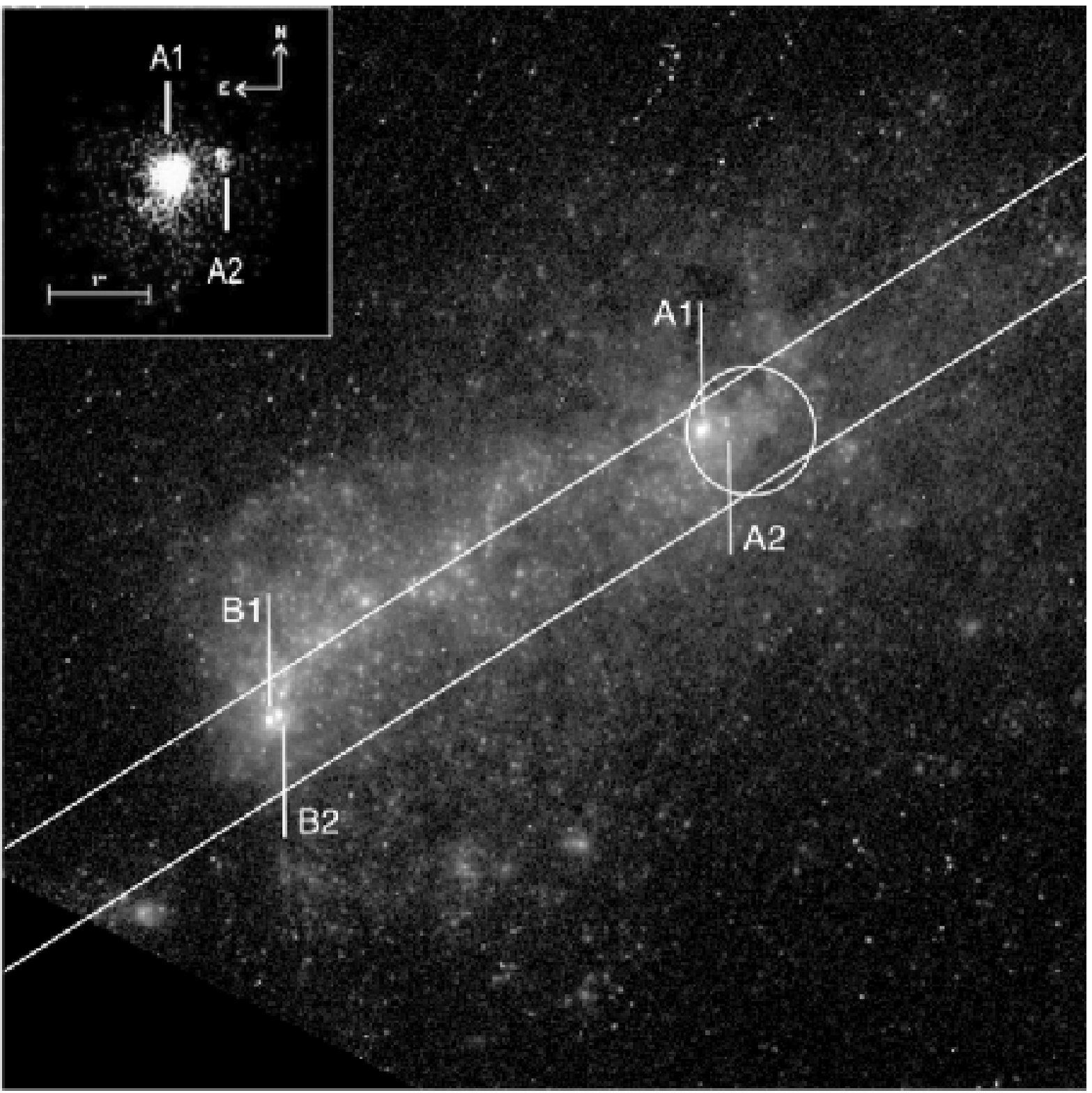}
\caption{Left.~False-color \hst~ACS HRC F330W/F555W/F658N (blue/green/red) composite image of NGC 3125 constructed from drizzled data with no background or continuum subtraction. North is up and east is to the left. The scale is logarithmic with blue/green/red scale limits of low=0/0/0 and high=1.5/5/2. Right.~Grayscale ACS HRC F330W image of NGC 3125. We overlay the footprints of the COS 2.5''-in-diameter aperture and STIS 52''x2'' long-slit. The inset on the upper left is the target acquisition confirmation image that was obtained with COS. It spans the wavelength range 1650-3200 \AA. We identify the positions of clusters A1, A2, B1, and B2 on the UV images. North is up and east is to the left.}
\label{fig1}
\end{figure}

NGC 3125 (Tol 3) is a nearby ($D=11.5$ Mpc\footnote{Value derived from the Galactic Standard of Rest velocity using a Hubble constant of Ho=75 km s$^{-1}$ Mpc$^{-1}$, adopted for comparison with \cite{sch99}, C04, and H06.}) blue compact dwarf galaxy of Large Magellanic Cloud (LMC) metallicity ($12+\rm{log\,[O/H]}= 8.3$, H06). The left panel of Figure~\ref{fig1} shows a composite NUV+optical image of NGC 3125 that we constructed from archival multidrizzled \hst~images with no further processing. The images were taken with the High-Resolution Channel (HRC) of the Advanced Camera for Surveys (ACS) as part of program 10400 (PI Chandar). The galaxy is dominated by two emission regions, NGC 3125-A and NGC 3125-B \citep{vac92}. As in H06 we subdivide region A into UV bright and highly extinguished components A1 and A2, respectively; and region B into components B1 and B2. In the right panel of Figure~\ref{fig1}, we show the ACS F330W image of NGC 3125, the locations of the above components, the STIS and COS aperture footprints, and the COS target acquisition (TA) NUV image of region A as an inset. Previous optical ground-based spectroscopy shows that WR stars are present in components A1 and A2, but it is unclear if they are present in B1 and/or B2 (\citealt{sch99}, hereafter S99; H06). Hereafter, we refer to these regions, which are knots of star formation, as clusters.

\subsection{New STIS Observations \& Data Reduction}

As part of program \hst~GO 12172 (PI Leitherer), we used STIS for obtaining co-spatial spectroscopy over the wavelength range from 1150 to 10270 \AA~of the two principal emitting regions in NGC 3125, i.e., A and B. For this purpose we used the $2$-wide long-slit; CCD, NUV-MAMA, and FUV-MAMA detectors; and G140L, G230L, G430L, and G750L gratings. The MAMA and CCD detectors have fields of view (FOVs) of $25\times25$ and  $51\times51$, respectively, and plate scales of 24.6 and 50.8 mas per pix in imaging mode, respectively. As shown in the right panel of Figure~\ref{fig1}, regions A and B both fit within the long-slit. Prior to taking the STIS spectra, we used the CCD detector and F28x50LP aperture to obtain an acquisition image. We used this image to check the orientation of the STIS spectra. Table~\ref{tab1} gives the wavelength coverages, nominal spatial and spectral resolutions (nominal), and start, stop, and exposure times of the different STIS exposures. The data were processed by the CALSTIS pipeline, which includes the following steps: conversion from highres to lowres pixels (MAMA), linearity correction, dark subtraction, cosmic-ray rejection (CCD), combination of cr-split images (CCD), flat fielding, geometric distortion correction, wavelength calibration, and photometric calibration.

\begin{figure}
\rotate
\epsscale{1}
\plotone{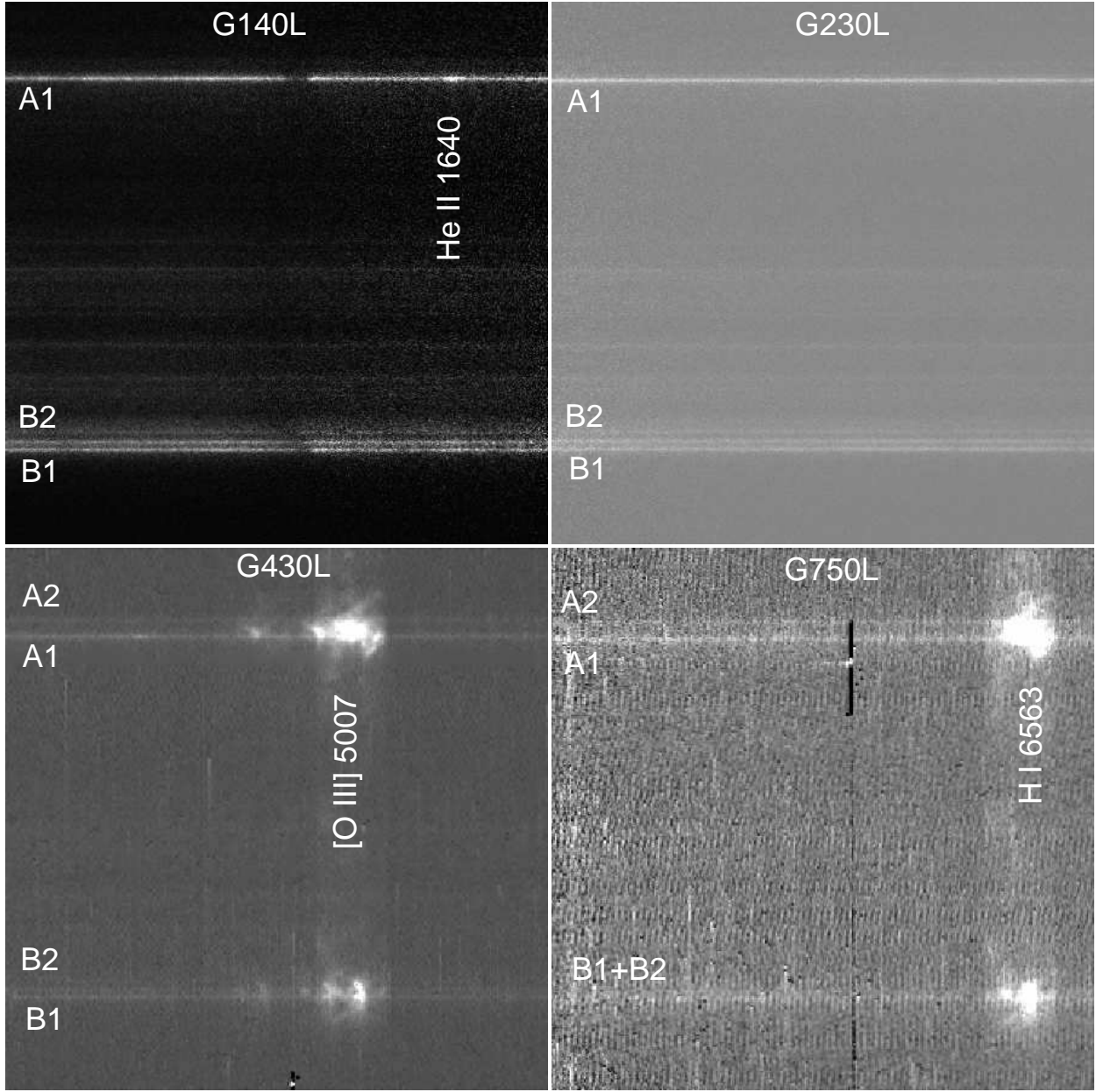}
\caption{Portions of the STIS G140L, G230L, G430L, and G750L 2D spectra. We identify the spectral traces of clusters A1, A2, B1, and B2, and the positions of prominent emission lines. The traces of A2 are faint in the G140L and G230L images due to reddening. The traces of B2 are faint in the G430L and G750L data.}
\label{fig2}
\end{figure}

\begin{figure}
\plotone{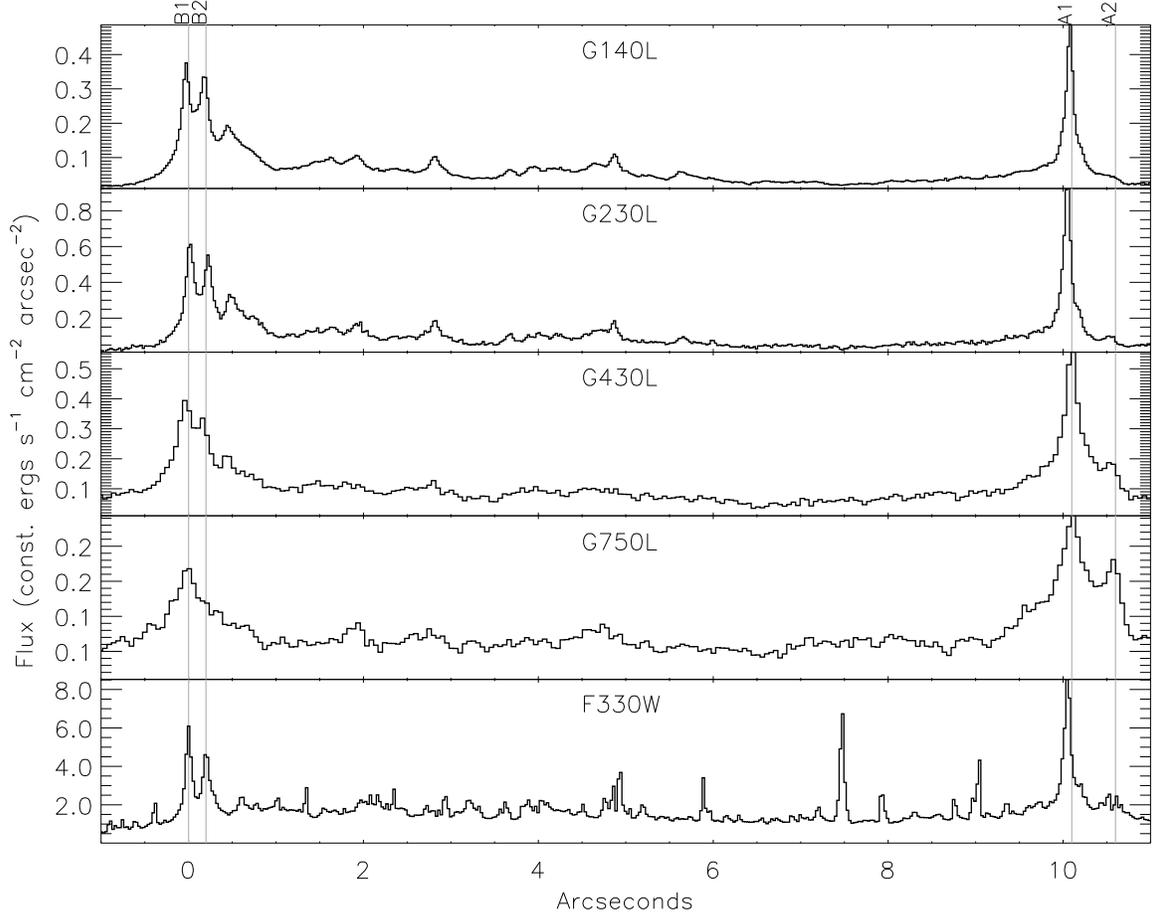}
\caption{First four panels from the top. Column plots extracted from continuum regions of the STIS 2D spectral images.  The width of the column is along the dispersion direction and equal to 100 pixels. The flux  is in units of 10$^{-12}$ erg s$^{-1}$ cm$^{-2}$ arcsec$^{-2}$). The bottom axis shows position (arcsec) along the cross-dispersion direction. The plate scales corresponding to the top two/next two panels are 0.0246/0.0508 arcsec pixel$^{-1}$, respectively. Bottom panel. Column plot extracted from the direct ACS HRC F330W image. The flux is in units of 10$^{-14}$ erg s$^{-1}$ cm$^{-2}$ arcsec$^{-2}$. The width of the column is 100\% along the x-axis. We mark the positions of clusters A1, A2, B1, and B2 with vertical lines which are labeled on the top panel. The figure shows the correspondence between spectral traces and clusters, that B1 and B2 merge to form one peak as they become fainter in the G750L panel, and that A2 becomes brighter towards longer wavelengths.}
\label{fig3}
\end{figure}

As there were no summed observations, we extracted one-dimensional (1D) spectra for clusters A1, B1, and B2 from the x2d  (MAMA) and sx2 (CCD) files, which contain two-dimensional (2D) spectral images.  Portions of the 2D spectral images are reproduced in Figure~\ref{fig2}. The correspondance between the spectral traces and clusters is shown in Figure~\ref{fig3}, where we display column plots extracted from continuum regions in the STIS 2D spectra. The width of the extracted columns is along the dispersion direction. As can be seen in figures~\ref{fig2} and~\ref{fig3}, A2 is highly obscured, particularly in the UV, and the contrast of B2 against the background is lowest at the highest wavelengths. No 1D spectrum was extracted for A2, since the data for this cluster is not useful for studying the massive star distribution. In addition, only a UV spectrum was extracted for B2. We used ATV \citep{bar01} to extract diffuse-background-corrected spectra from spectral orders 3 (G140L), 3 (G230L), 4 (G430L), and 3 (G750L). The fluxes were corrected for slit loss according to the data header keyword diff2pt. The scaling factors are 0.071, 0.064, 0.116, and 0.133 for G140L, G230L, G430L, and G750L, respectively. CALSTIS does not correct the CCD data for hot pixels and charge transfer efficiency (CTE). No CTE correction was performed as the signal to noise of the data was too low. We eliminated the hot pixels using a custom IDL sigma-filter routine and adopting $\sigma=7$.

\begin{figure}
\rotate
\epsscale{0.9}
\plotone{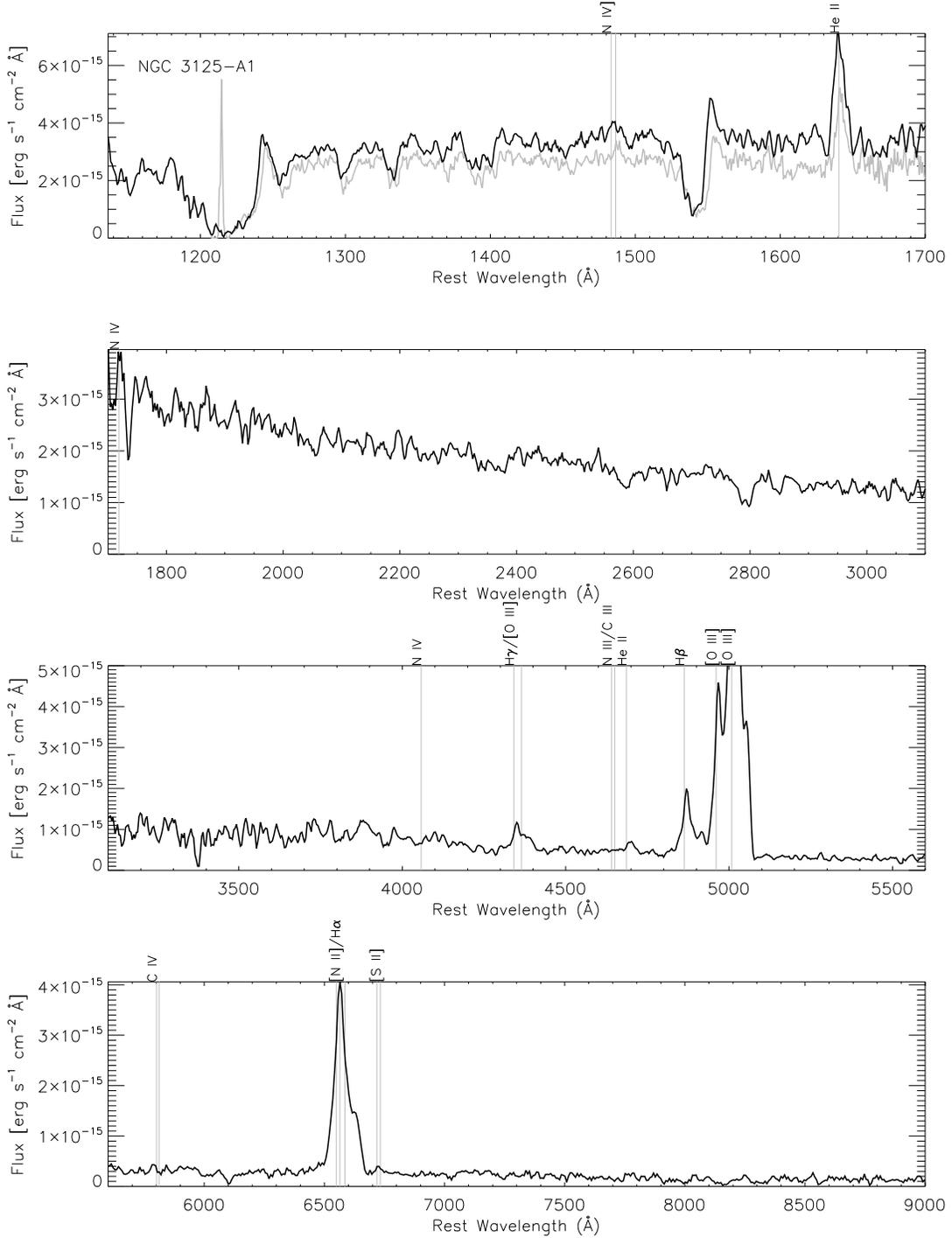}
\caption{Combined STIS FUV-to-NIR 25''x2'' long-slit spectrum of NGC 3125-A1(black curve). For comparison, we show the STIS FUV  25''x0.2'' long-slit spectrum of C04 (gray curve). The individual spectra are binned to the nominal spectral resolution of the grating with which they were taken and smoothed with a boxcar of n=5 pixels. All spectra are foreground-reddening corrected. We mark the rest-frame positions of the principal WR star and nebular features with vertical gray lines.}
\label{fig4}
\end{figure}

\begin{figure}
\rotate
\epsscale{0.9}
\plotone{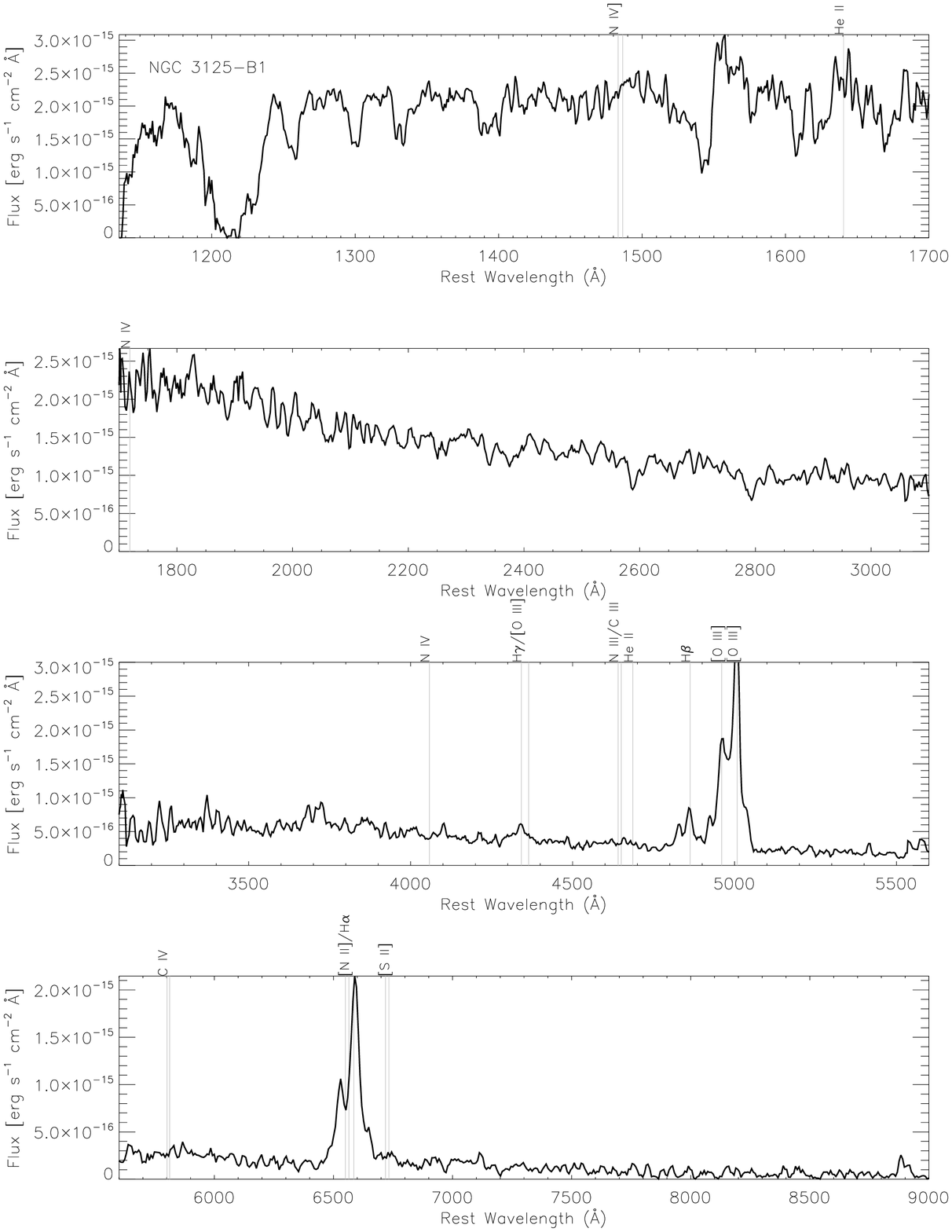}
\caption{Similar to Figure~\ref{fig4} but for NGC 3125-B1. There is no previous STIS spectrum for B1.}
\label{fig5}
\end{figure}

\begin{figure}
\rotate
\epsscale{0.9}
\plotone{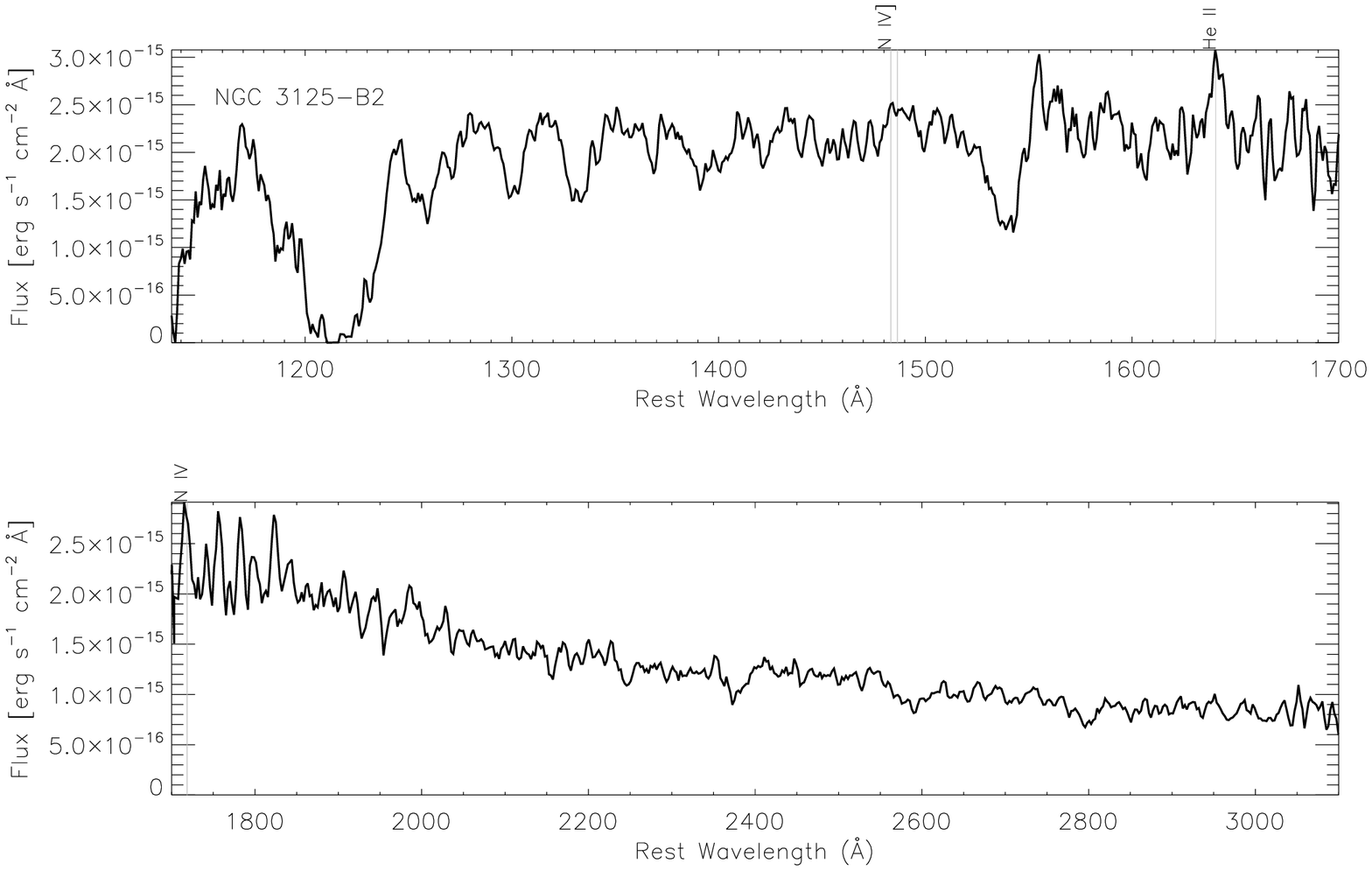}
\caption{Similar to Figure~\ref{fig4} but for NGC 3125-B2. There is no previous STIS spectrum for B2.}
\label{fig6}
\end{figure}

Further processing steps on the extracted 1D spectra include using linear interpolation to rebin to the nominal resolution of each grating  (the data are oversampled); adjusting the wavelength-scale zero points of clusters B1 and B2, using A1 as a reference (the zero point is completely dominated by the position of the clusters with the wide slit and we found no recognizable shift for A1, except for the redshift); correcting for foreground-reddening using $E(B-V)_G=0.076$ \citep{sch98} and the reddening law of \cite{mat90}; correcting for redshift using $z=0.003712$ or $v=1113$ km s$^{-1}$ (the latter is given by H\,{\sc\,i} 21-cm data); degrading data to the spectral resolution measured from contaminating Milky Way (MW) lines by smoothing the spectra with a boxcar and a window of n=5 pixels; and combining the spectra from the different gratings. The STIS spectra of clusters A1, B1, and B2 (uncorrected for intrinsic reddening) are shown in Figures~\ref{fig4},~\ref{fig5}, and~\ref{fig6}, respectively. We mark principal WR and nebular emission lines with vertical lines. For comparison, we overlay in Figure~\ref{fig4} the STIS spectrum of A1 published by C04. The older spectrum was obtained with a narrower 0.2''-wide slit, which explains its lower continuum flux. The reddening correction for dust within the emitting regions is discussed in the analysis section.

\subsection{New COS Observations \& Data Reduction}

As part of program \hst~GO 12172, we used COS for obtaining a high resolution spectrum of A1 covering the wavelength range from 1150 to 1450 \AA. For this purpose, we used the 2.5 in diameter primary science aperture (PSA), FUV-XDL detector (segments A and B), and G130M grating, whose nominal spectral resolution is 0.06 \AA~or 15 km s$^{-1}$ at 1200 \AA, i.e., an order of magnitude improvement over the archival STIS 25x0.2 long-slit presented in C04. The higher resolution is essential for our investigation of the spectral region around 1370 \AA, where a blend of Fe\vb~lines and the elusive O\vb~\lam1371 feature from the hottest most massive stars is located \citep{wal85}. We filled the 14.3 \AA~gap between segments A and B by using the two central wavelength settings 1300 and 1318 \AA~at focal plane position FP-POS=3. The right panel of Figure~\ref{fig1} shows the location and size of the COS footprint. The spectral observation was preceded by an image target acquisition, which was obtained with the NUV-MAMA detector. The TA confirmation image, which is shown as an inset in the right panel of Figure~\ref{fig1}, clearly shows the two components of region A. The last three rows of Table~\ref{tab1} give the wavelength coverage, nominal spatial and spectral resolutions, and start, stop and exposure times of the COS exposures. The data were processed through the first two steps of the CALCOS pipeline, i.e., (1) correction of the data for instrument effects (noise, thermal drifts, geometric distortions, pixel-to-pixel variations in sensitivity); and (2) construction of an exposure-specific wavelength-calibrated scale; and combined with the custom IDL co-addition routine described in \cite{dan10}. 

Further processing steps on the extracted 1D spectra include using linear interpolation to rebin to the nominal spectral resolution of the G130M grating (the data are oversampled); adjusting the wavelength-scale zero-point by 0.023 \AA~or $\sim$5.5 km s$^{-1}$ at 1250 \AA~(as this is the average residual between the observed and rest-frame contaminating geocoronal emission lines Ly$\alpha$ \lam1215.67 and O \ib~\lam\lam\lam1302.17, 1304.86, 1306.03 \AA); shifting spectra by an additional 0.06 \AA~(as this is the average residual between the observed and rest-frame contaminating MW absorption lines S\ii~\lam1250.61, S\ii~\lam1253.91, Si\ii~\lam1260.45, C\ii~\lam1334.62, and Si\iv~\lam1402.83); correcting for foreground-reddening using $E(B-V)_G=0.076$ \citep{sch98} and the reddening law of \cite{mat90}; and correcting for redshift. The effective spectral resolution given by the average $FWHM$ of the MW absorption lines is 0.3 \AA~or $\sim$75 km s$^{-1}$ at 1250 \AA. The intrinsic-reddening correction due to dust within A1 is discussed in the analysis section.

\section{DATA ANALYSIS}

\subsection{Intrinsic Reddening}

We aim to study the massive star distributions in A1, B1, and B2 via the WR-to-O-star number ratio. The number of O stars, $N(\rm{O})$, can be derived independently from the luminosity of the stellar continuum at 1500 \AA~or nebular H$\beta$~emission line.  The number of WR stars, $N(\rm{WR})$, can be derived independently from optical or UV WR lines. The $N(\rm{O})$ and $N(\rm{WR})$ values are sensitive to the de-reddening correction. We derived the intrinsic reddening, $E(B-V)_g$, using up to three independent methods, depending on the availability of the necessary data. The three methods, UV-slope, Balmer-ratio, and He\ii-ratio are described next.

\subsection{UV-Slope Method and Cluster Ages.}

To first order, the clusters under study can be considered to be single stellar populations (SSPs) and this is what we assume. The UV stellar continuum of an unextinguished instantaneous starburst can be modeled with a power law with spectral index $\beta$, such that the flux is given by $F\propto\lambda^\beta$ in the wavelength range 1250-2600 \AA~\citep{cal94}. Any deviation of the power-law exponent from the expected value for a young ($\le$10 Myr) starburst, i.e., $\beta=-2.6\pm0.2$, is assumed to be due to the effect of dust. The advantage of this method is that the reddening measurement is derived directly from the massive stars emission, not from the surrounding ionized gas, i.e., it represents the reddening of the \textit{stellar} continuum. In this method, $E(B-V)_g$ is the value providing the best match between the observed and expected spectral slopes. We measured the UV continuum slopes of the clusters over the wavelength range 1250-2600 \AA~using the STIS spectra. This method requires the determination of the cluster age and the selection of a reddening attenuation curve.  

The cluster ages were determined by comparing the observations to dust-free models, over the wavelength range 1200-1600 \AA. For this purpose, we divided each spectrum by a fit to the continuum. This avoids the reddening correction. We then followed the technique outlined in \cite{tre01} and adopted by C04 and H06, where the stellar wind N\vb~$\lambda\lambda$1238.8, 1242.8, Si\iv~$\lambda\lambda$1393.8, 1402.8, and C\iv~$\lambda\lambda$1548.2, 1550.8 resonance doublets (hereafter N\vb~1240, Si \iv~1400, and C\iv~1550) serve as an age chronometer. In summary, the P-Cygni profiles of the N\vb~and the C\iv~doublets decrease in strength as the O stars evolve and expire. On the other hand, the Si\iv~doublet develops a P-Cygni profile in giant and supergiant O stars. 

The SSP models were computed with the widely known package, Starburst99 \citep{lei99,lei10}, which is optimized to reproduce the average spectrophotometric properties of star-forming galaxies \citep{lei11}. Since the nebular oxygen abundance determined by H06 for NGC 3125 is similar to that of the LMC, we used the empirical LMC/SMC stellar spectral library for the models. Unfortunately, the latter library does not extend beyond 1600 \AA. The models were rebinned in order to match the effective spectral resolution of the UV observations, which is given by the FWHM of MW metal absorption lines. We adopted a \cite{kro01} IMF  with mass limits 0.1 and 100 M$_\odot$, a turnover stellar mass at 0.5 M$_\odot$, and lower/upper IMF exponents of $\alpha=$1.3/2.3. This stellar mass distribution is considered to be universally applicable at the high mass end \citep{bas10}. Finally, we used the Geneva stellar evolution tracks for non-rotating single stars with high-mass loss \citep{mey94}. Models that account for stellar rotation and close binarity are still under development \citep{eld08,lev12}.

\begin{figure}
\rotate
\epsscale{0.40}
\plotone{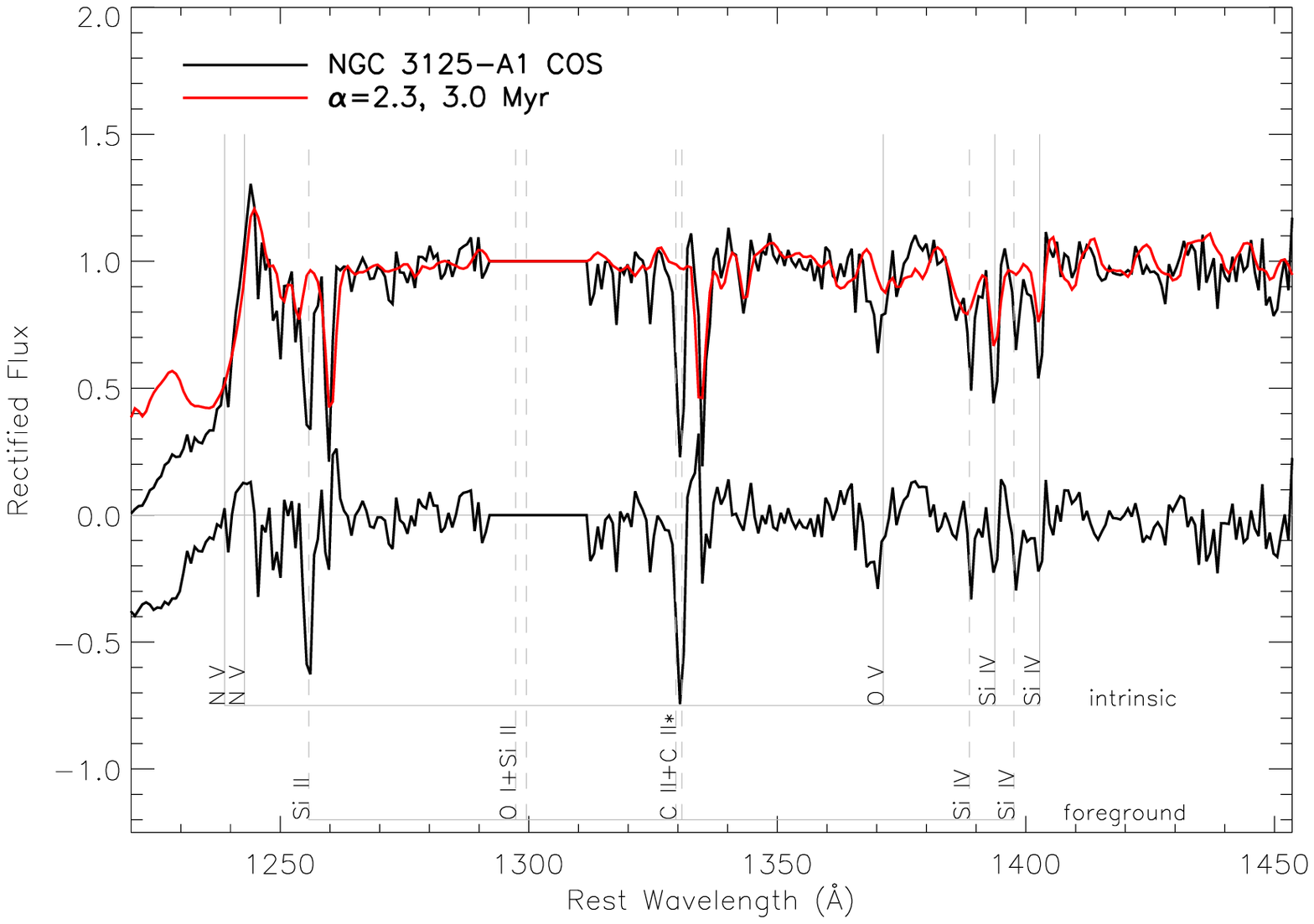}\plotone{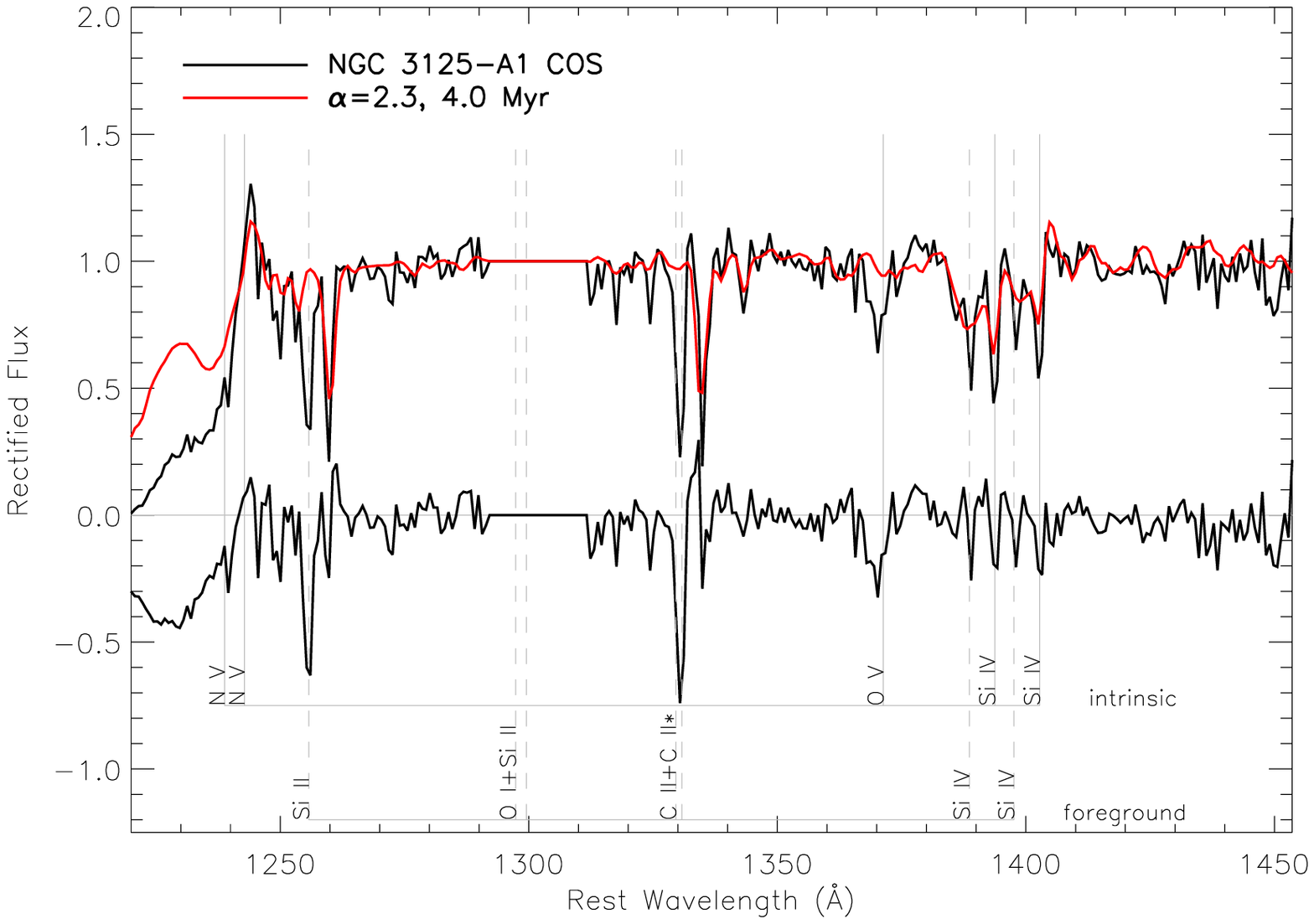}\\
\plotone{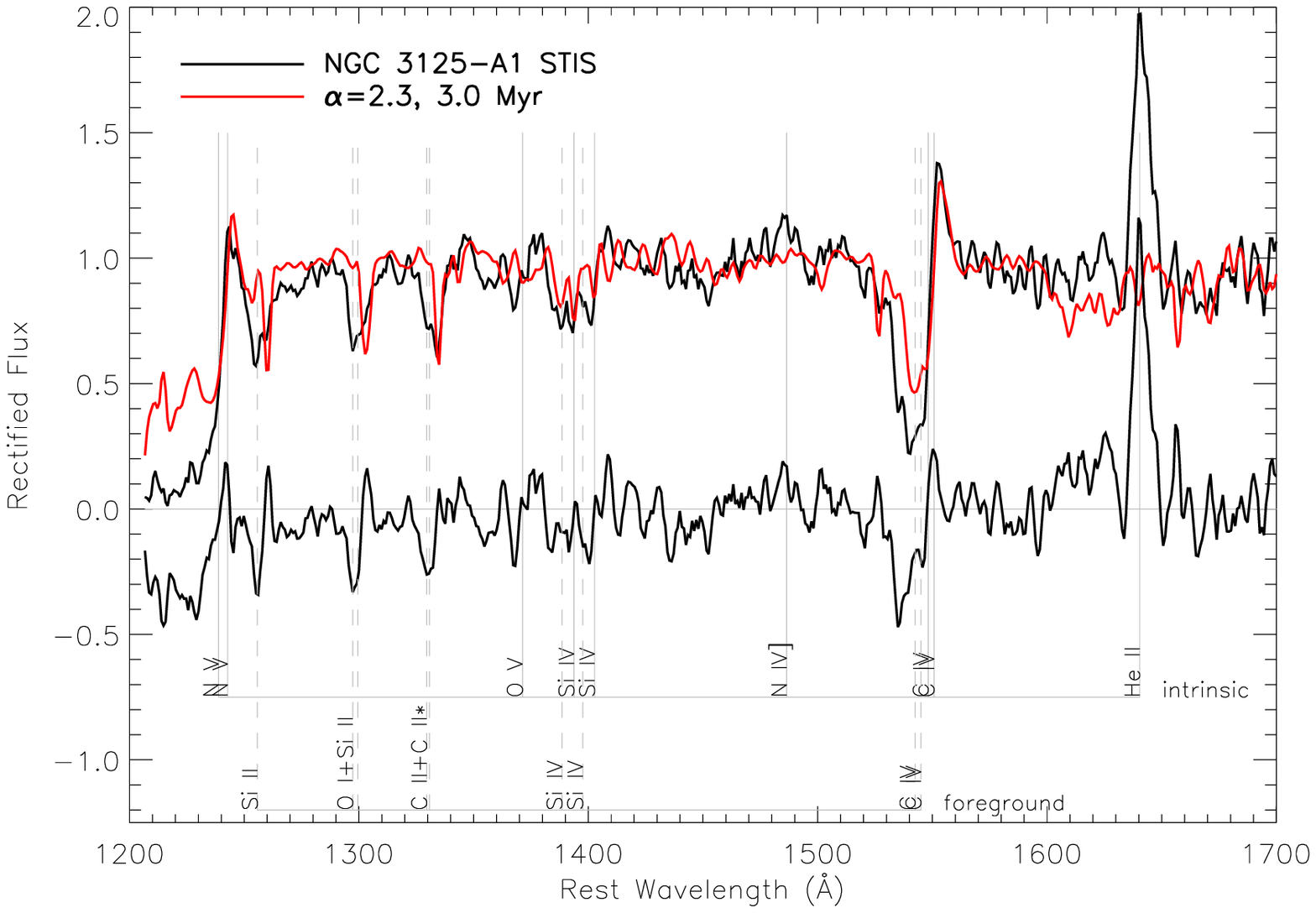}\plotone{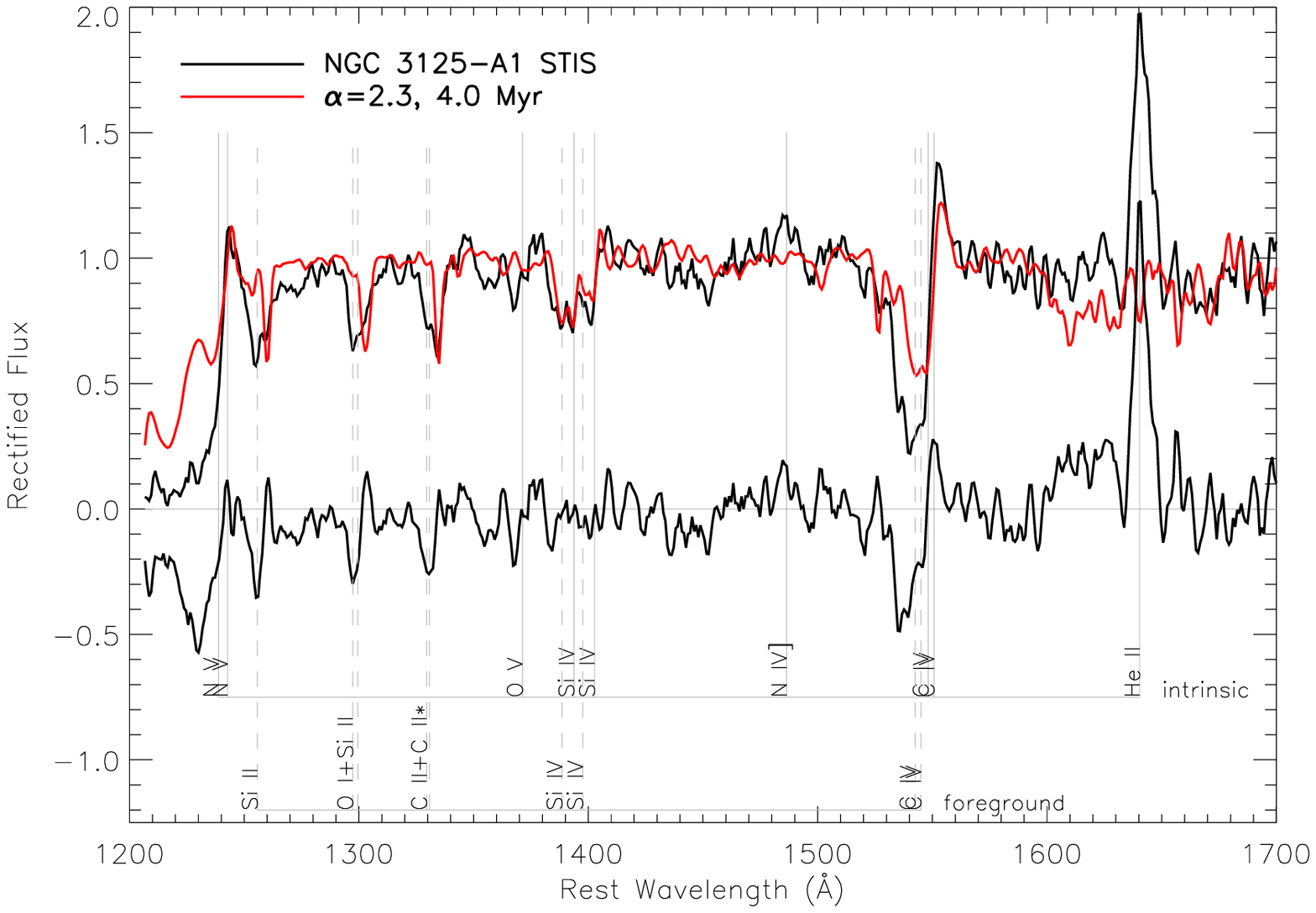}\\
\plotone{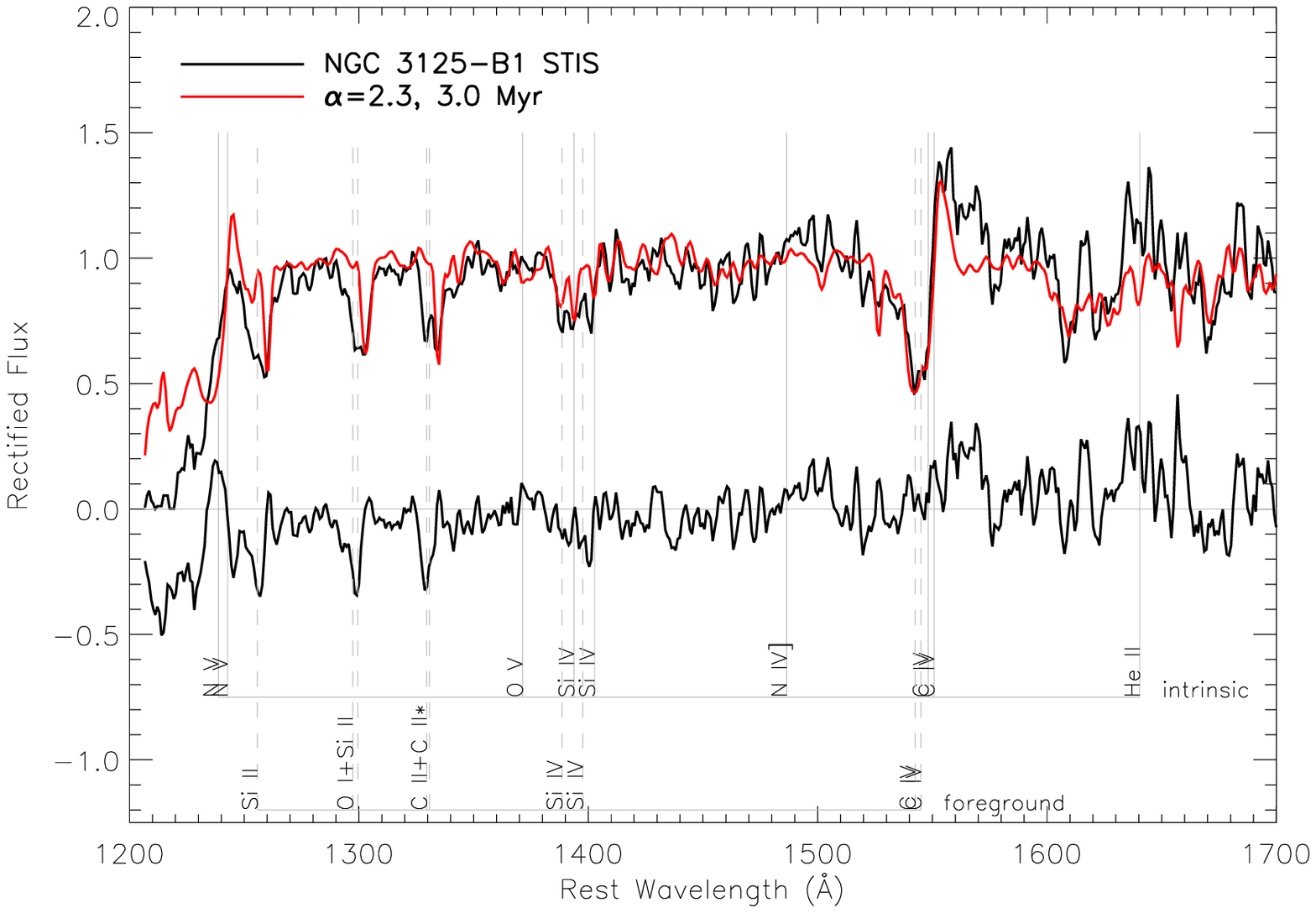}\plotone{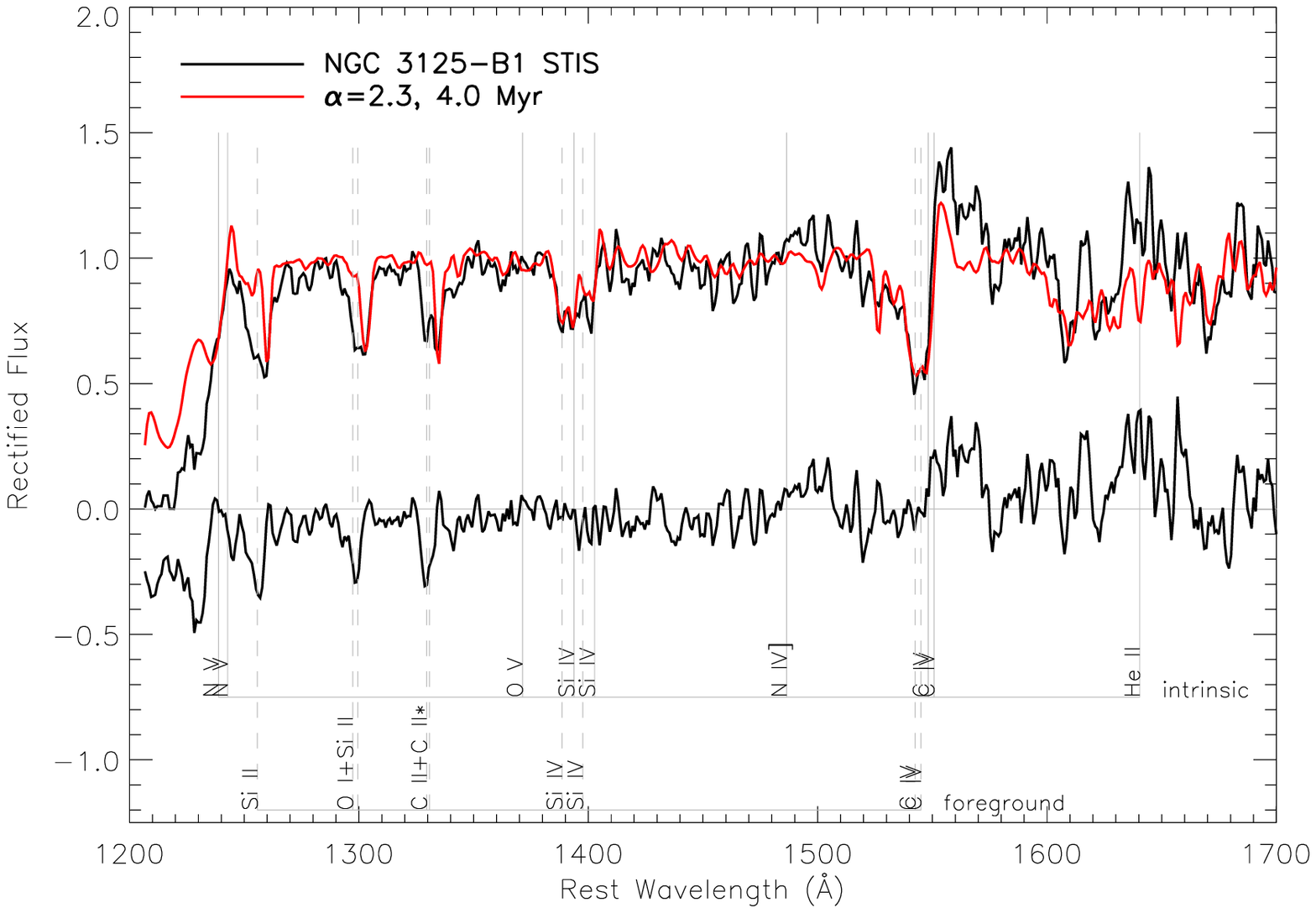}\\
\plotone{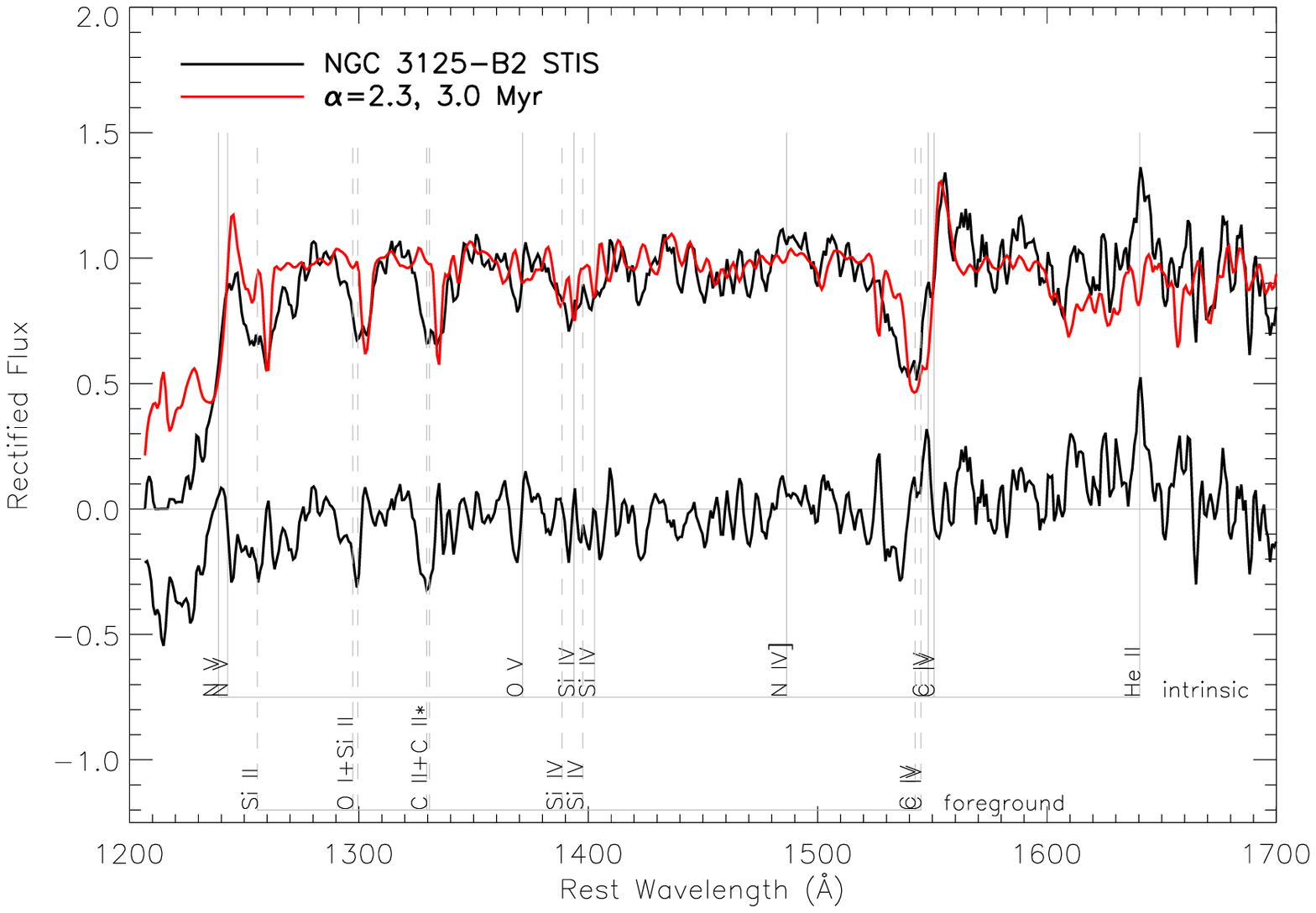}\plotone{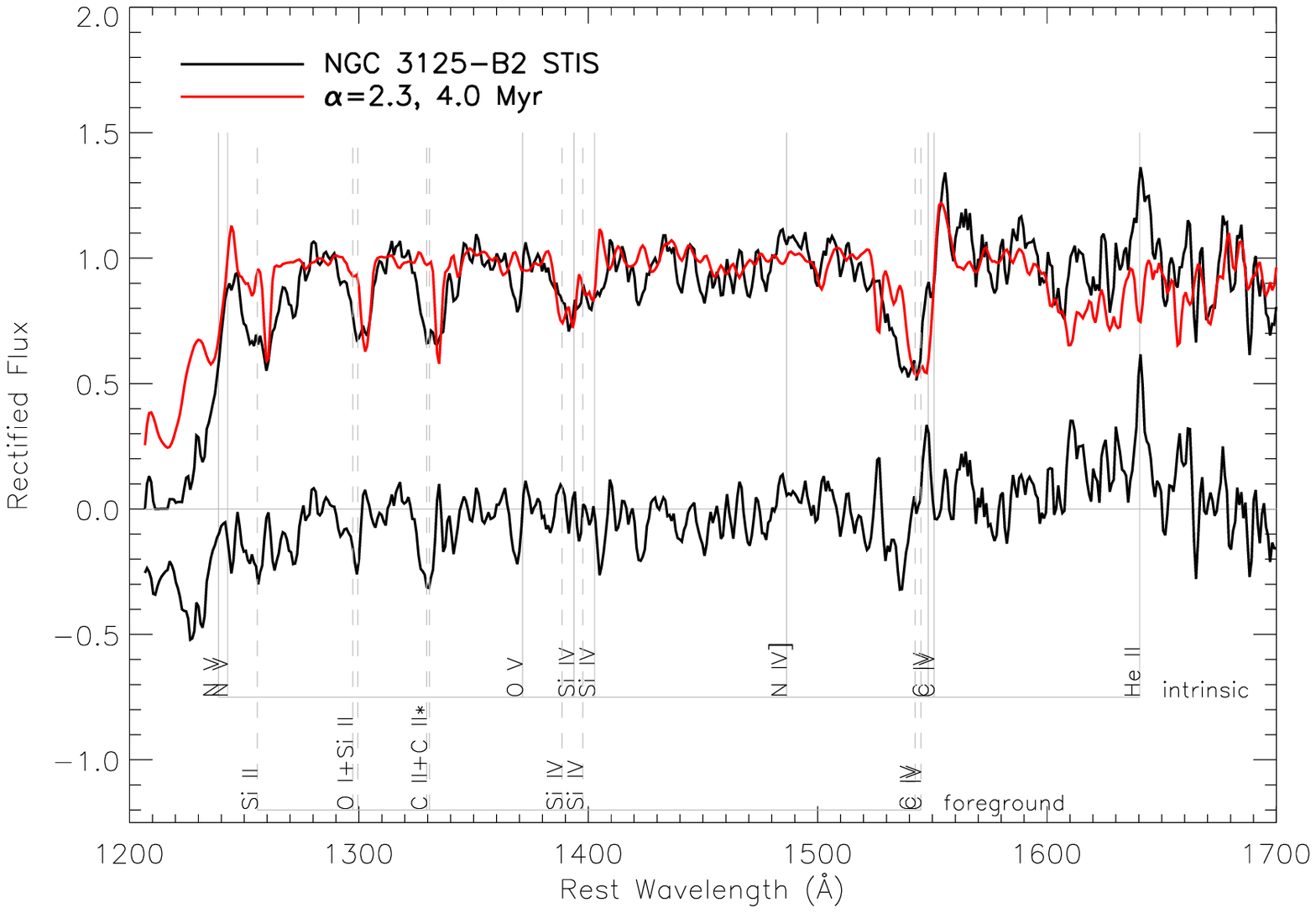}
\caption{Comparisons of COS and STIS observations (top black curves in each panel) with models (red curves). The models correspond to a \cite{kro01} IMF, ages of 3/4 Myr (left/right panels), and metallicities of $Z=0.004/0.020$ below/above 1600 \AA~(see text).  Row 1. COS observation of A1. Rows 2/3/4. STIS observations of A1/B1/B2. The observations are binned to the nominal spectral resolution of the corresponding grating, smoothed to match the spectral resolution of MW lines, and rectified. The spectral resolutions of the models and observations match. The lower curves in each panel are the residuals. We mark the rest-frame positions of dominant massive star features with vertical solid lines, and the blueshifted positions of some metal MW absorptions with vertical dashed lines.}

\label{fig7}
\end{figure}

The cluster ages were determined from the plots of Figure~\ref{fig7}. Since cluster A1 was also observed with COS at higher spectral resolution, for this cluster, we derived ages from the COS and STIS data. In Figure~\ref{fig7}, we mark the rest-frame positions of dominant intrinsic lines from massive stars with vertical solid lines. We also mark the blueshifted positions of some MW absorptions with vertical dashed lines. The latter produce large residuals because they are in the rest-frame of NGC-3125. The discrepancy between models and observations at $\sim1216$ \AA~is due to the H I Lyman-alpha MW absorption trough, which depends on Galactic latitude. When using the COS data of A1, the 3 Myr model best fits N\vb~1240, but the 4 Myr model best fits Si \iv~1400. In particular, we bring the reader's attention to the absorption components of N\vb~1240 and emission components of Si \iv~1400. Unfortunately the COS spectrum does not include C\iv~1550. However, the N\vb~1240 and C\iv~1550 profiles have similar spectral evolutions. When using the STIS data of A1, the 3 Myr model best fits C\iv~1550 but the 4 Myr model best fits Si \iv 1400. This is also true for clusters B1 and B2. In summary, the COS and STIS spectra yield similar results for A1, and the three clusters have similar ages. In addition, for A1 our ages are consistent with the results of C04 ($3\pm1$ Myr) and H06 (4 Myr), which are based on the older STIS spectrum published by C04. For knot B1+B2, H06 determined an age of 4 Myr based on H$\alpha$+K$_s$ photometry, which is consistent with our results for B1 and B2. Most importantly, the models do not reproduce the He\ii~$\lambda$1640~emissions and O\vb~\lam1371 absorptions that are present in clusters A1 and B2. For the rest of the analysis, we adopt an age of 3 Myr for the three clusters. Given the young ages derived for  A1, B1, and B2, it is reasonable to assume that any deviation of $\beta$ from the expected value for a young SSP is due to the effect of dust and not to age.

For the purpose of studying the reddening given by the UV continuum slope, we generated SSP models using the fully theoretical stellar library at $Z=0.004$ of \cite{lei10}, which provides continuous coverage from 1250-2600 \AA, unlike the empirical stellar library mentioned above. We did not use the fully theoretical stellar library for age-dating the clusters because they produce strong N \v~1240 and weak Si \iv~1400 P-Cygni profiles compared to observations. However, for studying the continuum slope, this is not an issue.

\begin{figure}
\epsscale{1}
\plotone{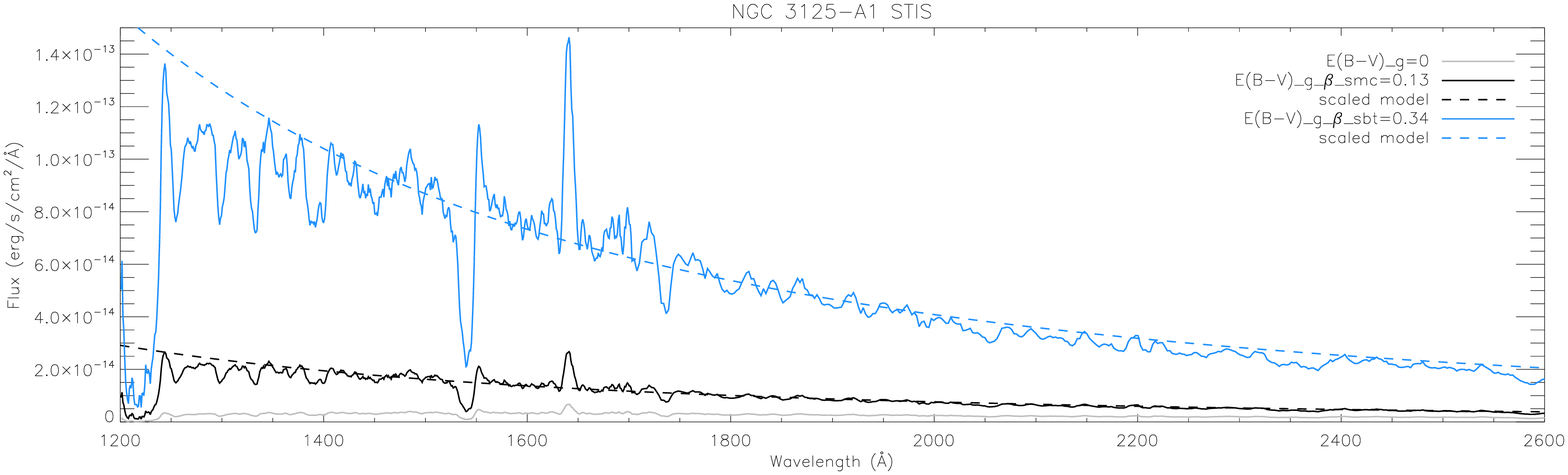}\\
\plotone{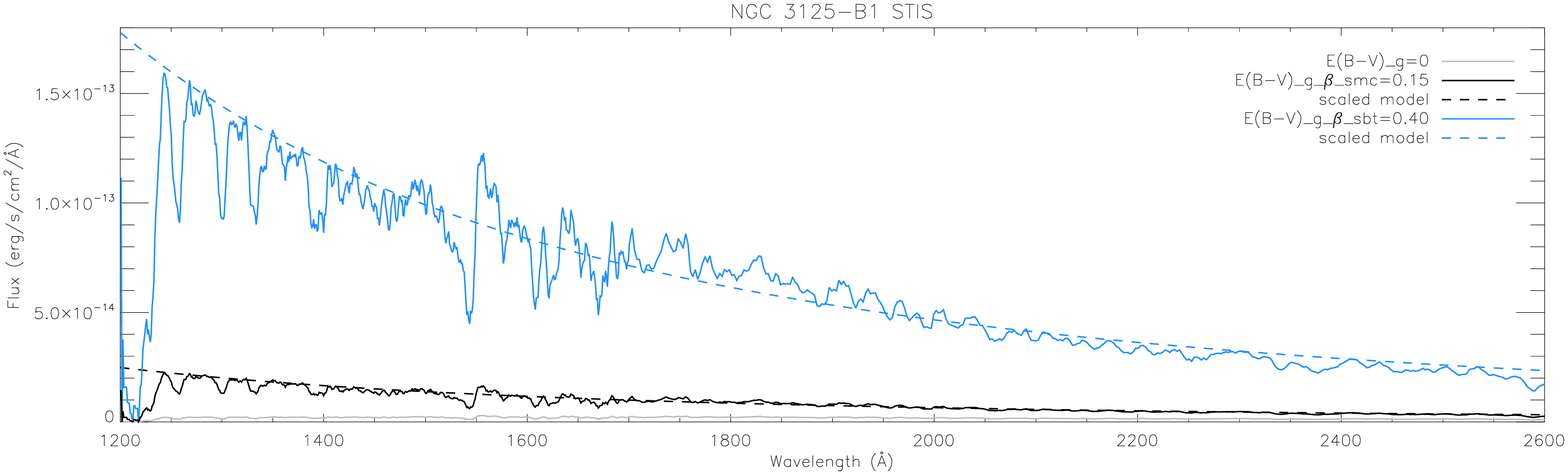}\\
\plotone{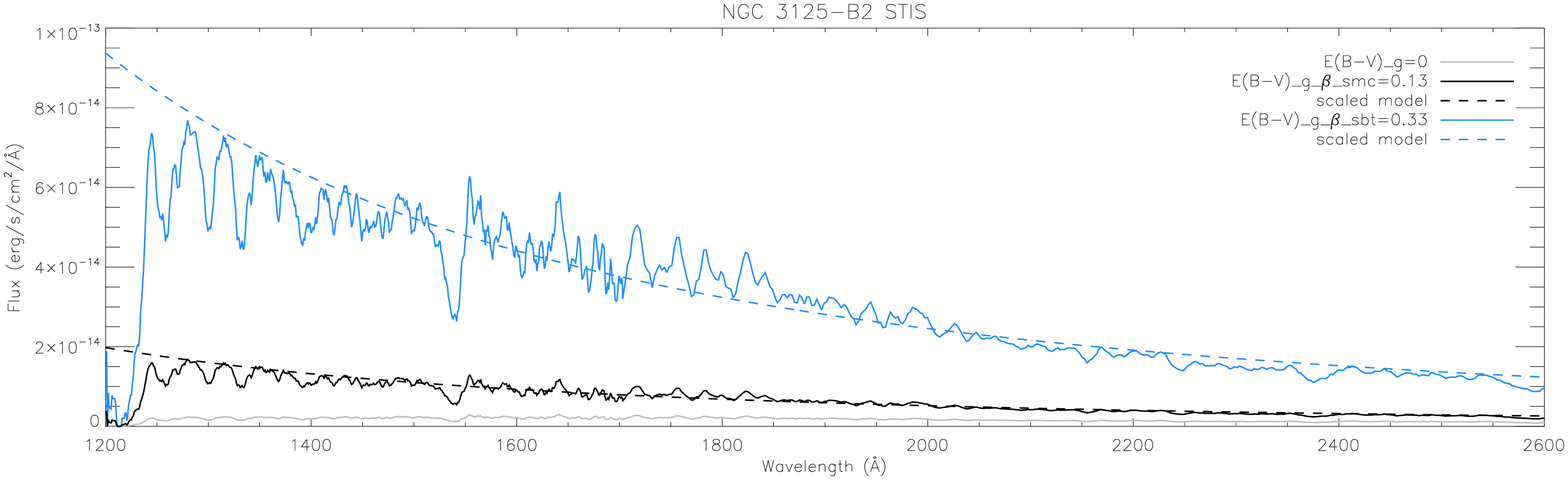}
\caption{STIS 1200-2600 \AA~spectra of NGC 3125-A1/B1/B2 (top/middle/bottom panels, respectively), corrected for
 intrinsic reddening using the SMC and starburst attenuation curves (solid black and blue curves respectively). For comparison, we also show the spectrum corrected for foreground reddening only (gray curve). The black and blue dotted lines are fits to the continuum of the same model, scaled to match the solid black and blue curves at 1900 \AA.}
\label{fig8}
\end{figure}

For de-reddening the spectra of the NGC 3125 clusters, H06 used the Small Magellanic Cloud (SMC) extinction curve because it provided the closest match to the complete UV spectral energy distribution for A1. On the other hand, C04 de-reddened the A1 spectra using the starburst attenuation curve of \citep{cal94}. We find that both curves reproduce well the spectral energy distribution of the clusters from 1250-2600 \AA. This can be seen in Figure~\ref{fig8}, where we compare the reddening-corrected spectra of clusters A1, B1, and B2 (top-to-bottom panels respectively) with a 3 Myr model. The model has been scaled to match the continuum flux at 1900 \AA. In Table~\ref{tab3} we give the $E(B-V)_{\rm{g,\beta,smc}}$ and  $E(B-V)_{\rm{g,\beta,sbt}}$ values that we obtained using the two reddening curves\footnote{we use the updated SMC extinction cuve of \citep{gor03}} as well as the$\beta$-slopes measured from reddening-corrected spectra. Since both attenuation curves yield a good match between the reddening corrected observations and the dust-free model, hereafter, we provide quantities derived using both curves. Note however that given the poin-like appearance of the clusters, the SMC extinction law is preferred. Indeed, the starburst attenuation law is more appropriate for extended regions.
We attribute the difference between the model and the reddening-corrected observations in the wavelength range $1700-1900$ \AA~for clusters B1 and B2 to the change of grating and contamination by each other's light. $E(B-V)_{\rm{g,\beta,sbt}}$ is significantly larger than $E(B-V)_{\rm{g,\beta,smc}}$ for the three clusters, which is expected as expected, since the two reddening curves diverge in the UV.

\subsection{Balmer-Ratio Method}

\begin{figure}
\epsscale{0.3}
\plotone{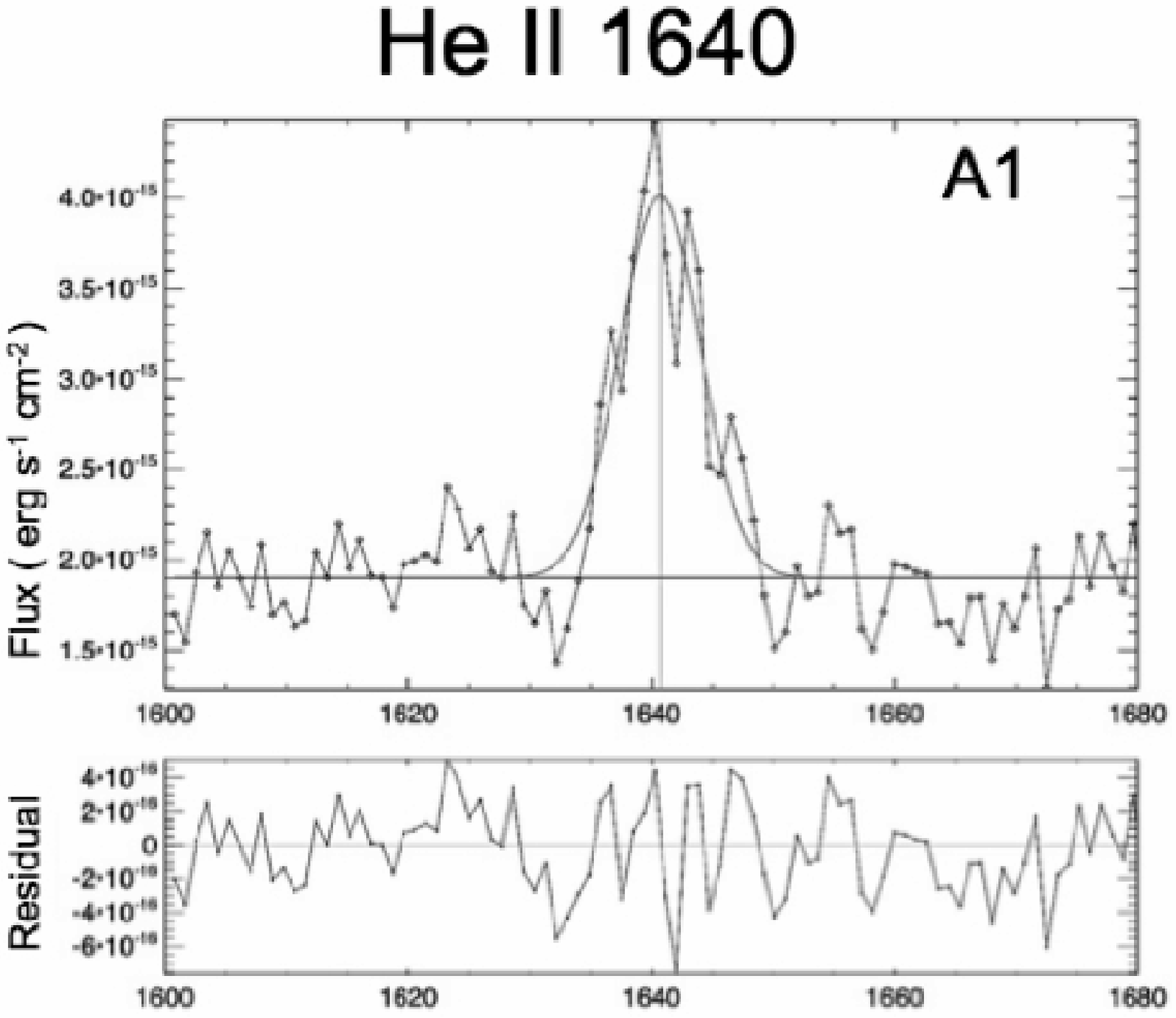}\plotone{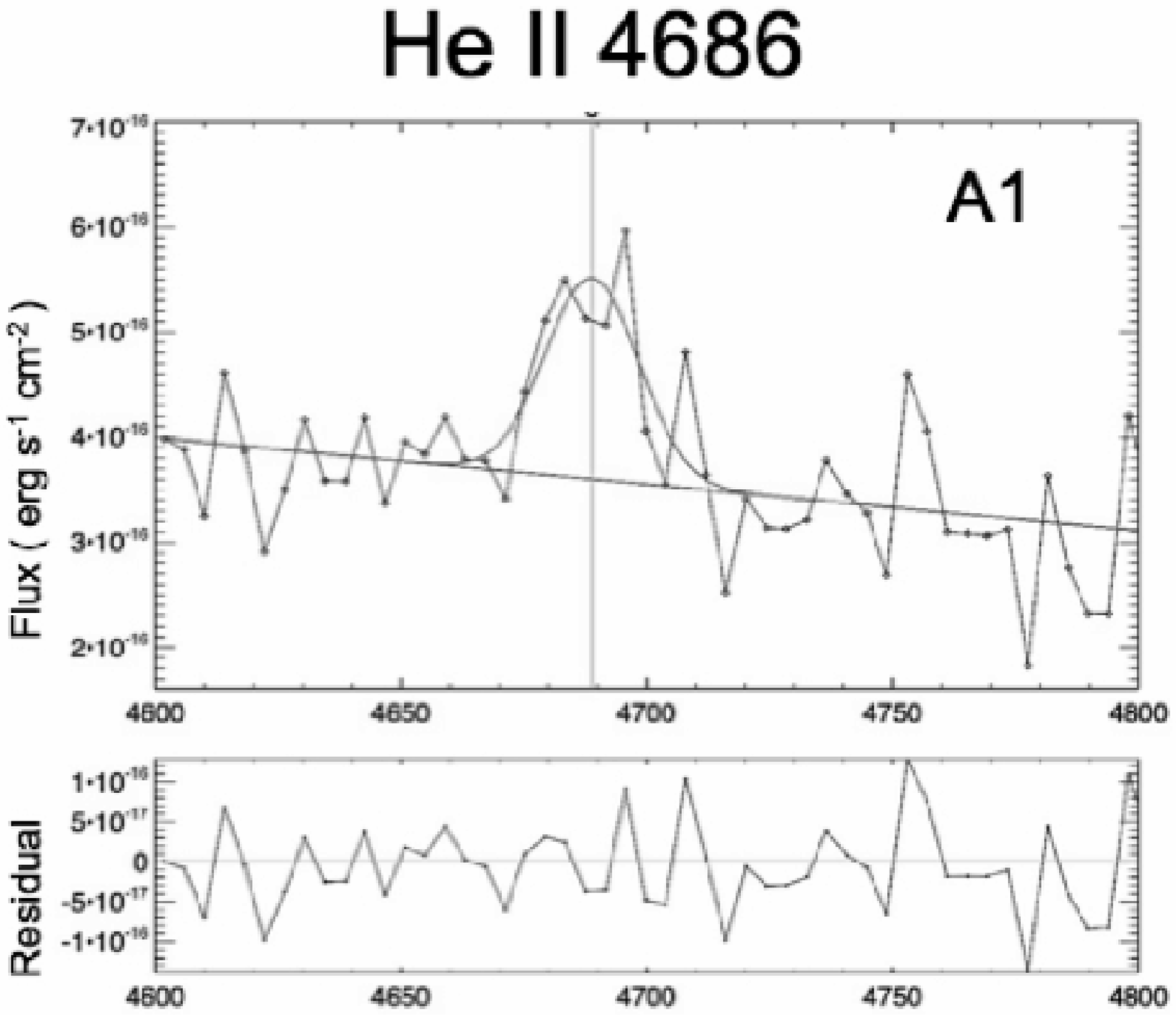}\plotone{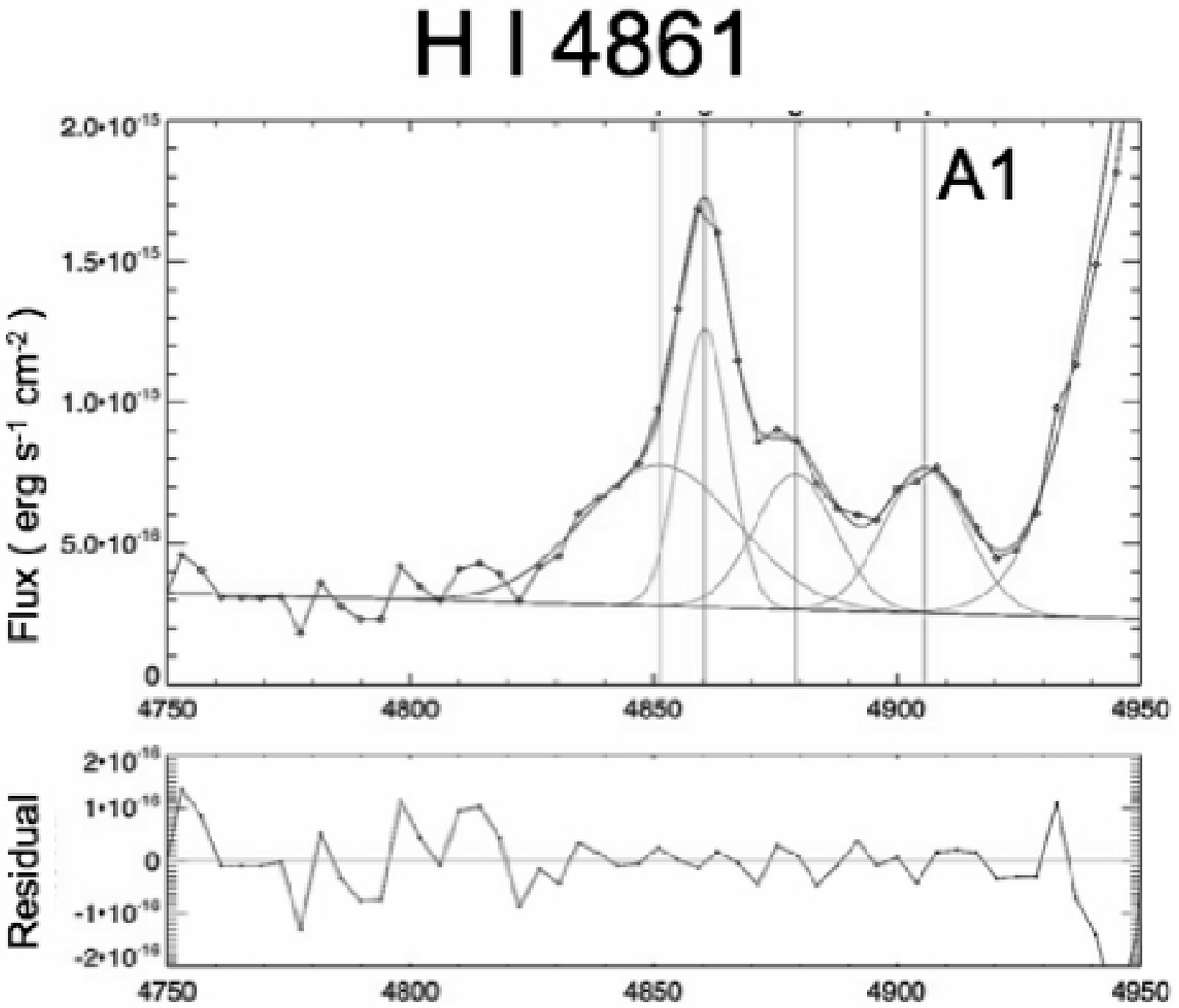}
\plotone{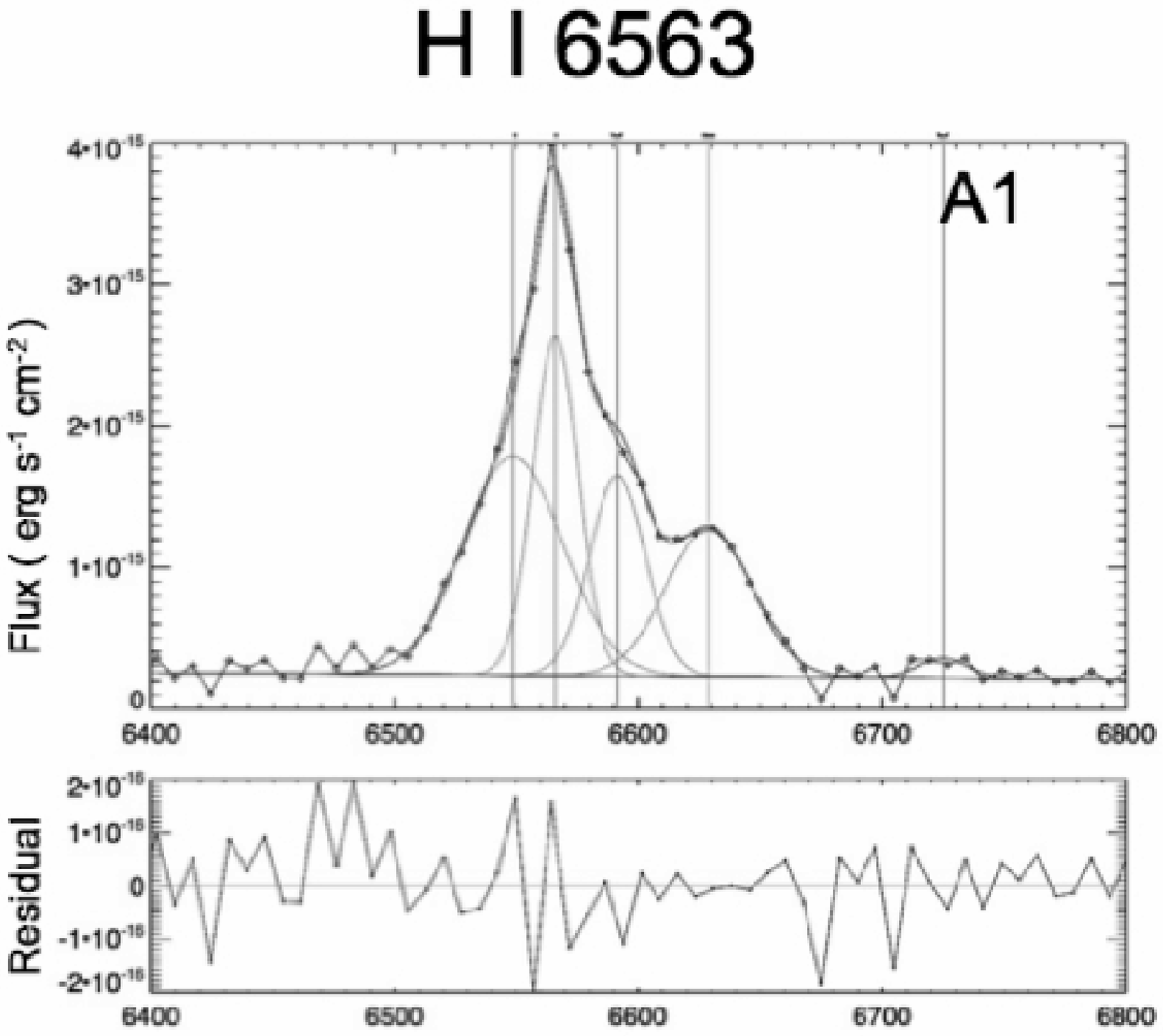}\plotone{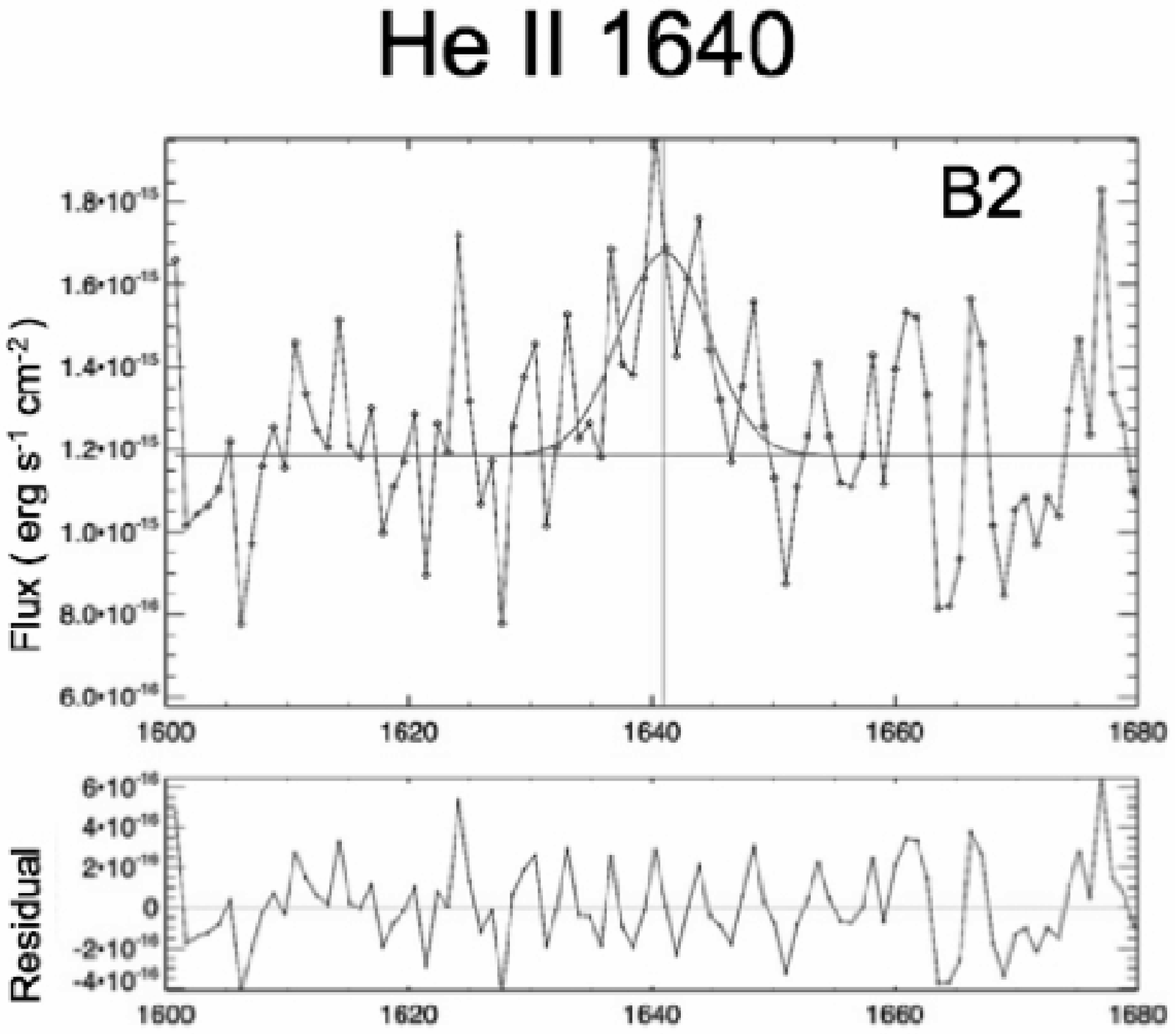}
\caption{Fits to emission lines from clusters A1 and B2. The Balmer lines have multiple components. We add these components to obtain the total H$\beta$ and H$\alpha$ fluxes.}
\label{fig9}
\end{figure}

Our optical spectra has lower signal to noise than that of H06 due to our lower exposure times. However, our nominal spectral resolution is better than their effective resolution, which is $FWHM=15$ \AA~for the VLT observations. Although the original purpose of our NUV-optical data was to study the reddening derived from the continuum slope, we decided to also measure the optical line fluxes of cluster A1 for comparison with results derived from the UV and by H06. The H$\alpha$ and H$\beta$ lines in A1 have multiple components. By adding the different components we obtained line fluxes that are in reasonable agreement with H06. We used the Balmer line ratio, H$\alpha$/H$\beta$, to derive the interstellar reddening in A1, assuming case B recombination theory for an electron density of 10$^2$ cm$^{-3}$ and a temperature of 10$^4$ K. This yields an expected ratio of 2.86 \citep{ost06}. We do not account for contamination of H$\alpha$ with the unresolved [N\ii]~\lam\lam6548, 6583 lines. However, these are expected to only contribute a few percent to the flux (see table 2 in H06, which shows that H$\alpha$ is 40 (30) times more luminous than [N\ii]~\lam6583 in region NGC 3125-A (-B)). Following H06, we corrected the Balmer lines for underlying stellar absorption from early-type stars, which is estimated to have an equivalent width of 2 \AA. This constitutes a negligible correction. Our fits to the Balmer lines were obtained with an IDL Gaussian-fitting routine and are shown in Figure~\ref{fig9}. In Table~\ref{tab3} we give the reddening-uncorrected Balmer-line fluxes, the $E(B-V)_{\rm{g,h,smc}}$ and  $E(B-V)_{\rm{g,h,sbt}}$ values derived from the Balmer-line ratio, and the Balmer-line ratios obtained from the reddening corrected line luminosities. We obtain similar $E(B-V)_{\rm{g,h}}$ values when using the starburst and SMC reddening curves. This is expected, since the two attenuation curves converge in the optical. The $E(B-V)_{\rm{g,h,smc}}$ and $E(B-V)_{\rm{g,h,sbt}}$ values are intermediate between $E(B-V)_{\rm{g,\beta,smc}}$ and  $E(B-V)_{\rm{g,\beta,sbt}}$, and larger than the value of $E(B-V)_{\rm{g,h,smc}}$ obtained by H06. Given the low quality of our optical data, our $E(B-V)_{\rm{g,h}}$ values are highly uncertain and should be viewed with care.

\subsection{He-Line Method}

For A1, we used a third method for deriving the intrinsic reddening, which compares the observed He\ii~\lam1640 to He\ii~\lam4686 line flux ratio to the expected value. In principle, this ratio provides a robust reddening measurement as it is independent of age since the two He\ii~lines arise in exactly the same short-lived stars. In addition, in contrast to the Balmer lines which measure the nebular reddening, the He\ii~lines measure the \textit{stellar} reddening. \cite{sch98} obtained 7.55 for a small sample of WN stars while \cite{cro06} obtained a higher ratio of 9.9 for a larger sample of LMC WN stars and 9-10 from theoretical WR models. We use an expected ratio of 9.9 in our reddening calculation. In Table~\ref{tab3} we give the reddening-uncorrected He\ii~line fluxes, and the reddenings obtained with the He-line method, i.e., $E(B-V)_{\rm{g,he,smc}}$ and  $E(B-V)_{\rm{g,he,sbt}}$. The fits to the He\ii~lines are shown in Figure~\ref{fig9}. We find the lowest reddening obtained so far, i.e.,  $E(B-V)_{\rm{g,he,smc}}=0.10$.

\subsection{Cluster Mass and Number of O Stars}

We derived the cluster mass, $M_{\rm{*}}$, by comparing the reddening-corrected stellar continuum luminosity at 1500 \AA, $L_{1500}$, to the expected value, $L_{1500,\,the}$, for a SSP of 3 Myr, $M_{\rm{*}}=10^6$ M$\odot$, $Z=0.004$, and \cite{kro01} IMF. To a first approximation, $M_{\rm{*}}=(10^6$ M$\odot)\times (L_{1500}/L_{1500,\,the})$.   We computed independent UV and optical estimates of the number of O stars in each cluster. These estimates are $N(O)_{1500}$ and $N(O)_{H\beta}$, respectively. We derived $N(O)_{1500}$ from $N(O)_{1500}=N(O)_{the}\times(L_{1500,\,the}/L_{1500})$, where we use $N(O)_{the}=2793$ and $L_{1500,\,the}=1.54\times10^{39}$, as predicted by the above model. Similarly, we derived $N(O)_{H\beta}$ from $N(O)_{H\beta}=N(O)_{the}\times(L_{4861,\,the}/L_{4681})$, where $L_{4861,\,the}=10^{40.193}$ erg s$^{-1}$, and $L_{4681}$ is the observed (reddening-corrected) H$\beta$ luminosity.  In Table~\ref{tab4} we give the values of $L_{1500}$, $L_{4861}$, and $M_{\rm{*}}$, and in Table~\ref{tab5} we give the values of $N(O)_{1500}$ and $N(O)_{H\beta}$. As can be seen in Table~\ref{tab5}, $N(O)$ can vary significantly depending on the reddening correction. $N(O)_{H\beta}$ is lower than $N(O)_{1500}$ by factors of 2-3. This could be due to our uncertain H$\beta$ fluxes, to the fact that the H$\beta$ emission extends beyond our slit, and to the fact that the H II gas is optically thin. For A1 H06 derived a much higher number of O stars from the optical imaging than the UV spectroscopy. Their explanation for this is that NGC 3125 hosts optically obscured young massive clusters which is supported by their VLT/ISAAC K-band imaging of the region. Our $N(O)_{1500}$ value for A1 is 467 versus 550 for H06. We find a higher $N(O)_{1500}$ for B1+B2 (630) compared to H06 (450). The differences in $N(O)$ values reflect the uncertainties in the measurements and methods. In our case, the expected $N(O)_{the}$ is only as certain as the stellar evolution tracks are. In addition, if the cluster IMF is flatter than Kroupa, then $N(O)_{the}$ and $L_{1500,\,the}$ would be underestimated, and so would $N(O)_{1500}$.

\subsection{Number of WR Stars}

The number of WR stars, $N(\rm{WR})$, is the sum of the WN, WC, and WO WR subtypes, which are defined in \cite{vac92}. The numbers $N(\rm{WN})$, $N(\rm{WC})$, and $N(\rm{WO})$ are obtained by dividing the luminosities of emission lines from populations of each subtype, by single star templates. Unfortunately, we do not detect the C\iv~\lam5808 emission from WC stars reported by S99 \& H06. This could be due to the low signal-to-noise of our optical data. We are only able to estimate the number of WN stars and assume that the line fluxes used to derive $N(\rm{WN})$ have no contribution from WC stars. Note that at an age of 3 Myr, our SSP model predicts a WC/WN ratio of zero.

WN stars show the products of CNO hydrogen burning and are dominated by helium and nitrogen emission lines, including N\iv~\lam1488, He\ii~\lam1640, N\vb~\lam1718, N\iv~\lam4058, N\vb~\lam\lam4604, 4620, N\iii~\lam\lam4634, 4640, and He\ii~\lam4686.  Thus, in particular, $N(\rm{WN})$ can be  estimated independently from He\ii~\lam1640 or He\ii~\lam4686. Numerical subtypes ranging from WN2 to WN9 are assigned to WN stars, such that strong He\ii~4686 emission plus the presence of the N\iii~\lam4640-C\iii~\lam4650 bump is indicative of a predominantly late WN population, while weak N\iv~\lam4058 suggests a dominant mid WN subtype. In addition, He\ii~4686 is narrow in the mid-WN (WN5-6) type and narrower in the late-WN (WNL) type \citep{smi96,cro06,mof89}.

In A1, S99+C04 report lines predominantly from WNL stars, while H06 report lines predominantly from WN5-6 stars. Establishing the dominant WN subtype, WNL or WN5-6, is beyond the scope of the present paper and we compute the number of WN5-6 stars. We measured line fluxes for He\ii~\lam1640 in A1 and B2, and for He\ii~\lam4686 in A1. Table~\ref{tab3} reports the reddening-uncorrected line fluxes and equivalent widths of these lines. The equivalent widths were corrected for the instrumental profile. Table~\ref{tab4} gives the corresponding reddening-corrected luminosities and Table~\ref{tab5} gives the values of $N(WN5-6)$ obtained with the different reddenings. When the same reddening is used for calculating $N(O)$ and $N(WN5-6)$, the $N(WN5-6)/N(O)$ ratio ranges from 0.19 to 0.56. Since $N(O)_{H\beta}$ is lower than $N(O)_{1500}$, $N(WN5-6)/N(O)$ is less when we use $N(O)_{H\beta}$ instead of $N(O)_{1500}$. If we use different reddenings for  $N(O)$ and $N(WN5-6)$, then we obtain $N(WN5-6)/N(O)$ as low as 0.05. When we use only ultraviolet data or only optical data, $N(WN5-6)/N(O)$ is independent of the choice of the attenuation curve. Our most robust ratios are $N(WN5-6)/N(O)_{\rm{1640,1500,\beta,\beta, smc}}=0.23$ for A1 and $N(WN5-6)/N(O)_{\rm{1640,1500,\beta,\beta, smc}}=0.10$ for B2, both of which are in good agreement with the ratios found by H06. Finally, for A1, the equivalent width of $EW_{1640}$ and $N(WN5-6)/N(O)$ are much larger than model predictions for a standard stellar population.

\section{ANALYSIS AND DISCUSSION OF COS DATA}

\subsection{O V 1371 \AA~Feature}

Central to our investigation is the spectral region around 1370 \AA, where a blend of Fe\vb~lines and the elusive O\vb~$\lambda$1371 feature are located. O\vb~is observed only in the very hottest, most massive stars \citep{wal85}, and it is absent in the composite spectrum of local star-forming galaxies published by \cite{lei11}. The line is seen as a very weak feature in a few starburst regions when spectra of the highest quality are available but no other starburst region displays a feature as strong as in NGC 3125-A1. However, complicating the interpretation is the contamination by photospheric Fe\vb~blends. Test calculations with Starburst99 models \citep{lei99} suggest enhanced strength of this feature for a very young starburst age and/or for an IMF with an excess of the most massive stars. Since this feature is potentially a new tracer of very massive stars, we obtained a COS/G130M medium resolution spectrum of A1. 

\begin{figure}
\epsscale{1}
\plotone{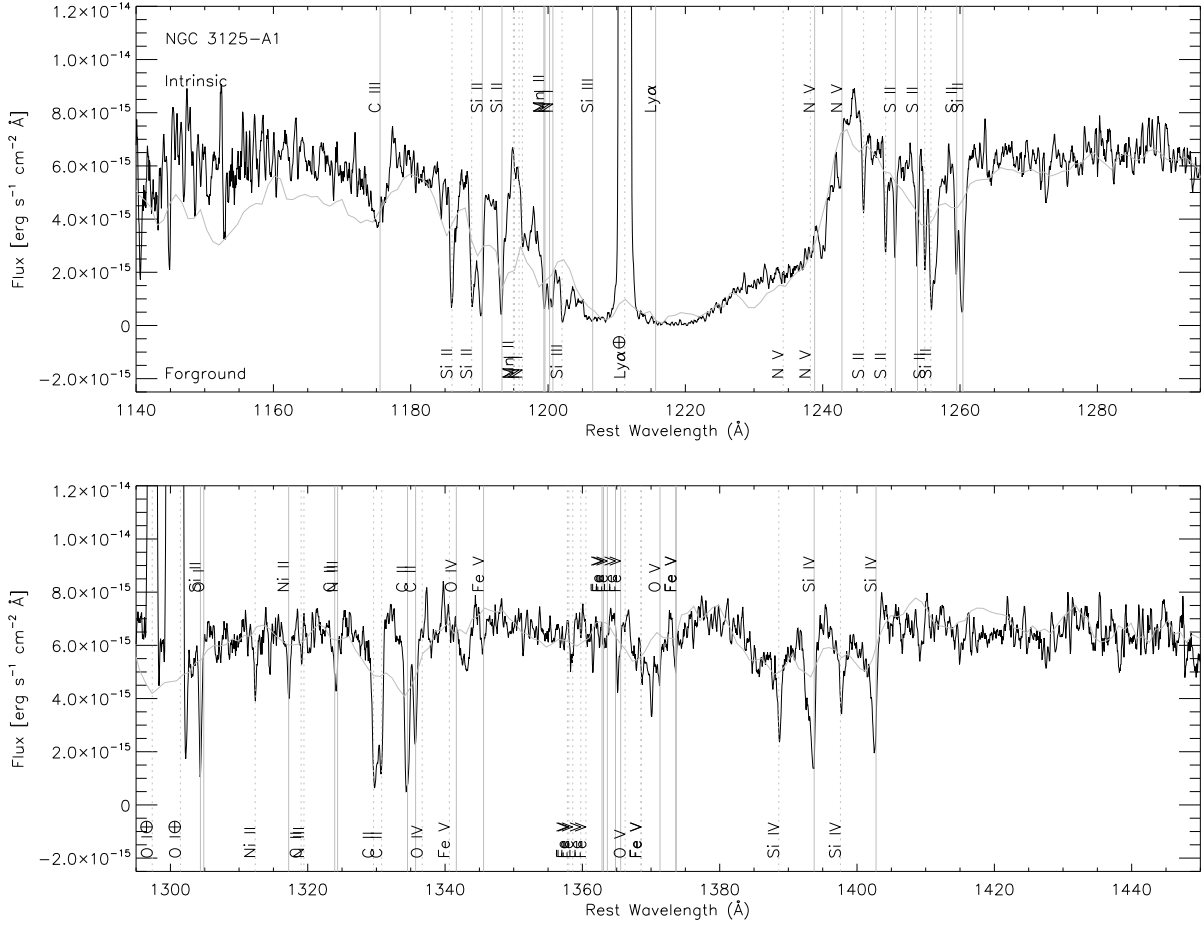}
\caption{COS/G130M spectrum of NGC 3125-A1 (black curve), and lower spectral resolution STIS/G140L spectrum (gray curve). We mark the rest-frame wavelengths of intrinsic lines at the top (vertical solid lines) and foreground lines at the bottom (vertical dotted lines). The foreground Ly$\alpha$ and O I emission lines are marked with a plus symbol because they are geocoronal lines. The higher spectral resolution of the COS spectrum is particularly useful for the study of  O\vb~$\lambda$1371 line.}
\label{fig10}
\end{figure}

The COS spectrum in shown in Figure~\ref{fig10}, where we have identified the positions of prominent foreground and intrinsic lines with dotted and solid lines respectively. The COS spectrum spans the rest-frame wavelength range, $\sim1140-1450$ \AA, and has an effective spectral resolution of $FWHM=0.3$ \AA. For comparison, we overlay the lower spectral resolution STIS/G140L spectrum in gray. The massive star features are more clearly seen in the COS spectrum.

\begin{figure}
\epsscale{0.48}
\plotone{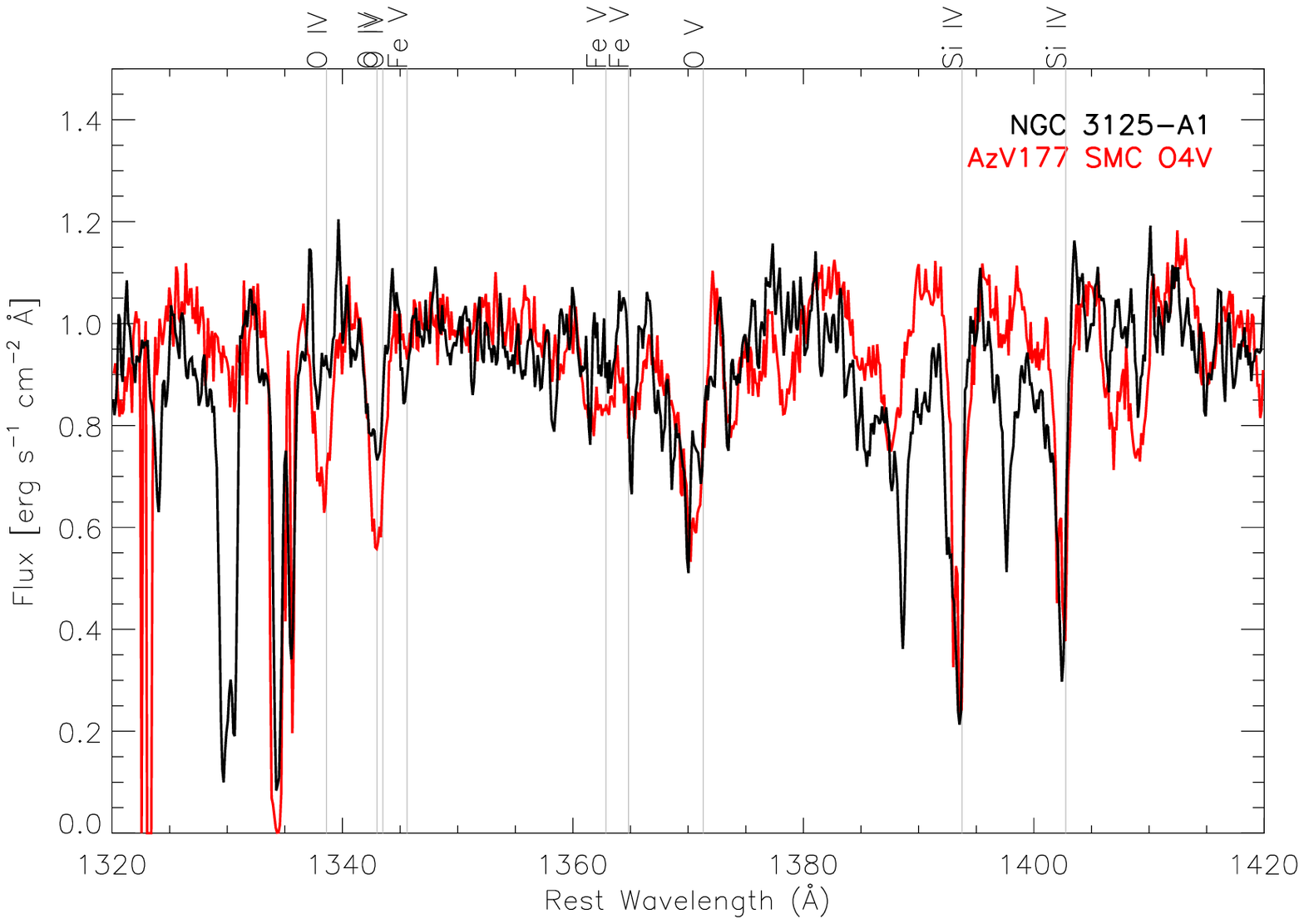}\plotone{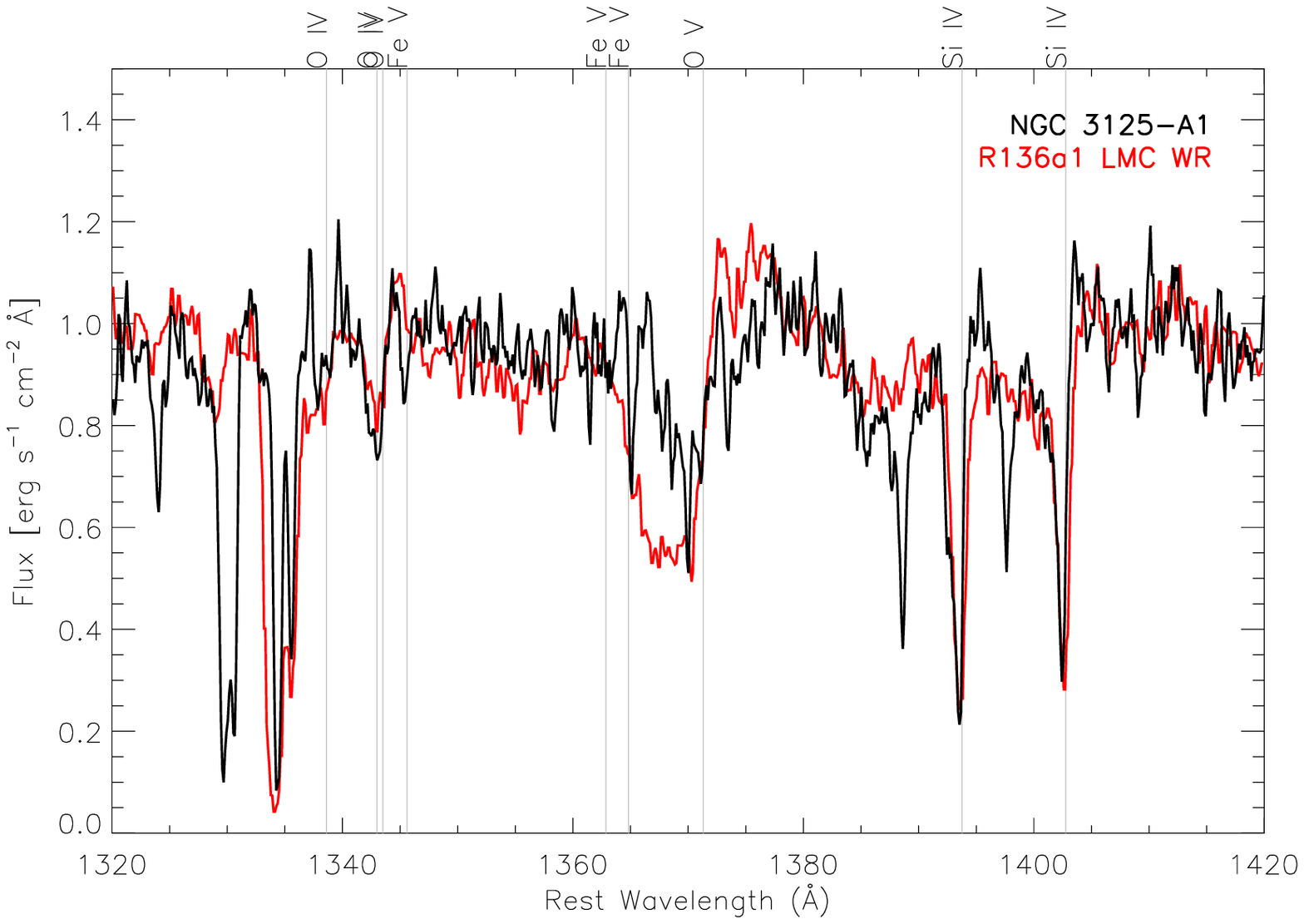}
\plotone{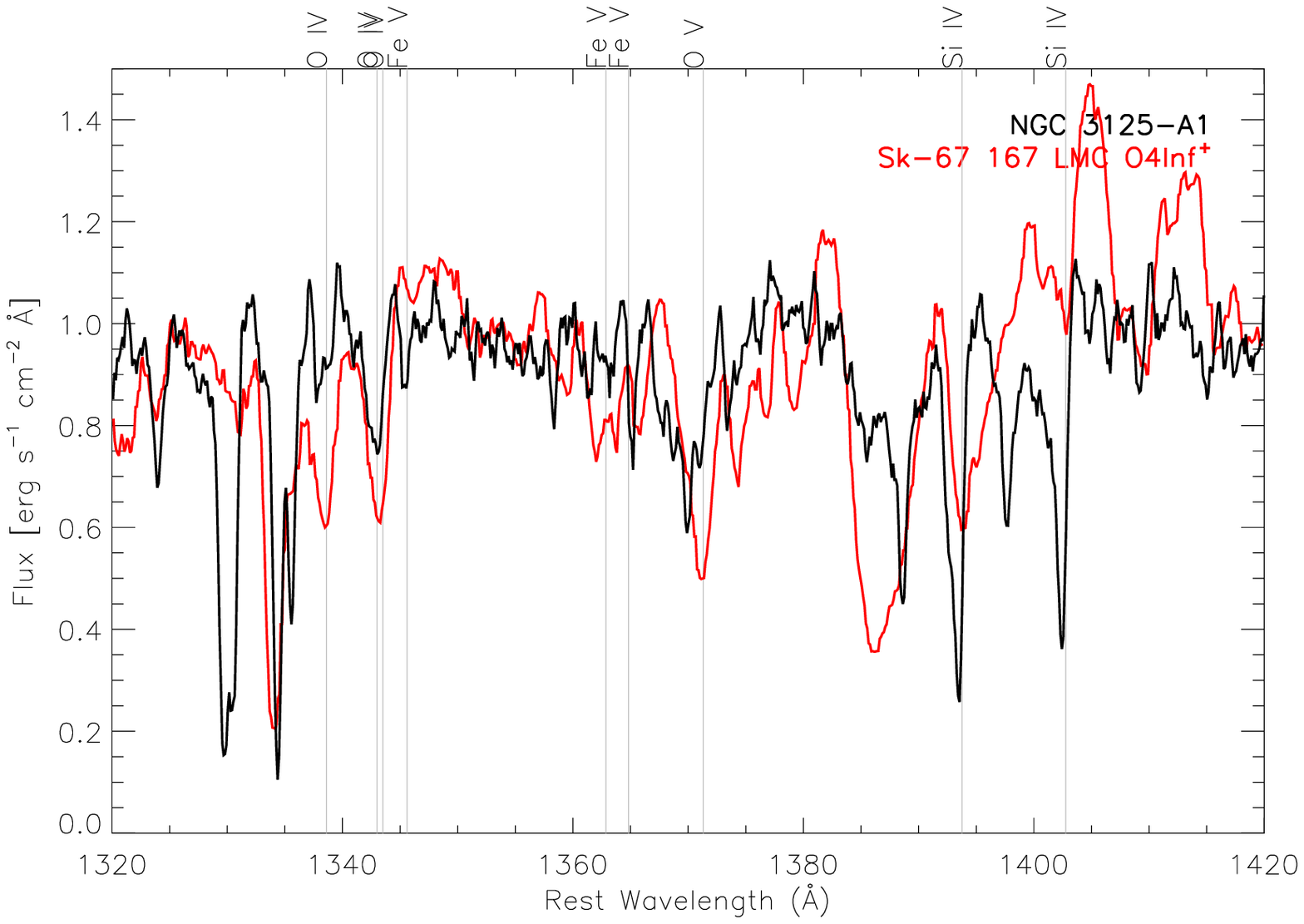}\plotone{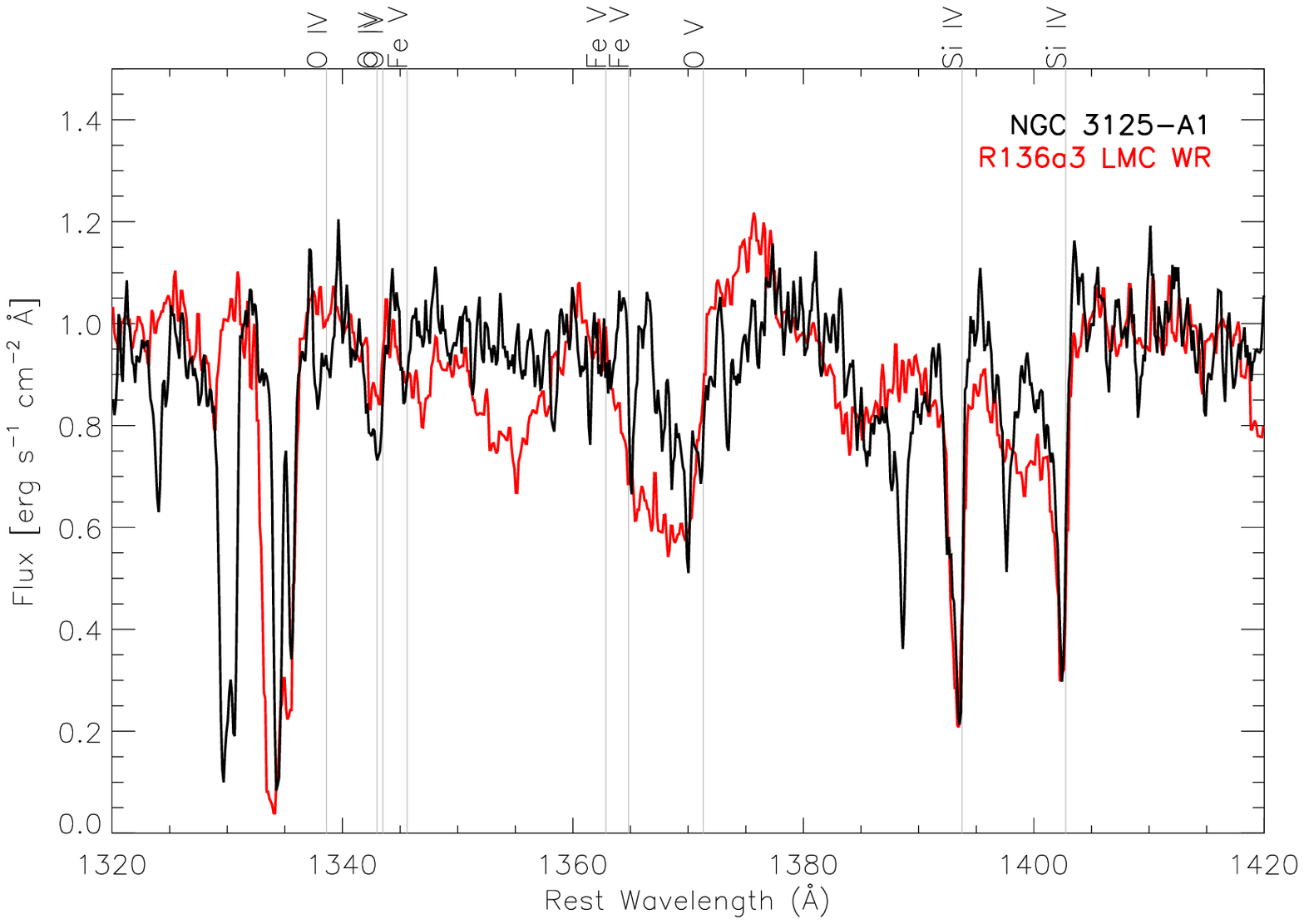}
\caption{Comparison of rectified COS/G130M spectrum of NGC 3125-A1 (black curves) with the spectra of individual LMC and SMC massive stars (red curves). We show the spectral region $1320-1420$ \AA. Top-left. \hst~COS/G130M spectrum of SMC O4 V star Azv 177 from \cite{bou13}. Bottom-left.  \hst~FOS/G130H spectrum of LMC O4 If stars Sk-67 167  from \cite{wal95}. Right-panels \hst~GHRS/G140L spectra of LMC WR stars R136a1 (top) and R136a3 (bottom) from \cite{dek97}. The spectral resolutions of NGC 3125-A1 and the stars were set to approximately match. We mark the rest-frame wavelength positions of some stellar lines discussed in the text with vertical lines.}
\label{fig11}
\end{figure}

Massive and hot O and WR stars are characterized by strong O\vb~\lam1371 absorption at SMC and LMC metallicities \citep{wal95,dek97,bou13}. In Figure~\ref{fig11} we compare the COS spectrum of NGC 3125-A1 to the spectra of individual LMC and SMC O4 V, O4 If, and WR stars. How well the individual stellar spectra compare to the spectrum of A1 depends on the line. In the figure, we mark the rest-frame wavelength postions of some stellar lines with vertical lines as examples. For O\vb~\lam1371, the agreement is better with the O4 V SMC star than with the LMC supergiant and WR stars. For Si\iv~1400 the agreement is good except with the supergiant. For the O\iv~lines, the agreement is better with the WR stars. These differences can be explained if there are more WR than O stars in A1. In any case, Figure~\ref{fig11} clearly shows that O\vb~\lam1371 from massive stars is detected in A1.

\section{DISCUSSION}

\subsection{Reddening Law and Massive Star Content}

The choice of the de-reddening method severely affects the massive star distributions derived from optical and UV spectroscopic observations. This must be considered when interpreting low and high redshift data, as differences as high as an order of magnitude in the number, $N(\rm{WN})$, derived from the optical and UV can arise just from this choice (Table~\ref{tab5}). Once the attenuation law has been selected, one must still consider differences in the intrinsic reddenings obtained with different methods. For A1, and using an SMC extinction law, we found that $E(B-V)_g= 0.10$ (He-lines), 0.13 (UV-slope), and 0.34 (Balmer-ratio).

\subsection{Interpretation of the Observations}

\begin{figure}
\epsscale{0.48}
\plotone{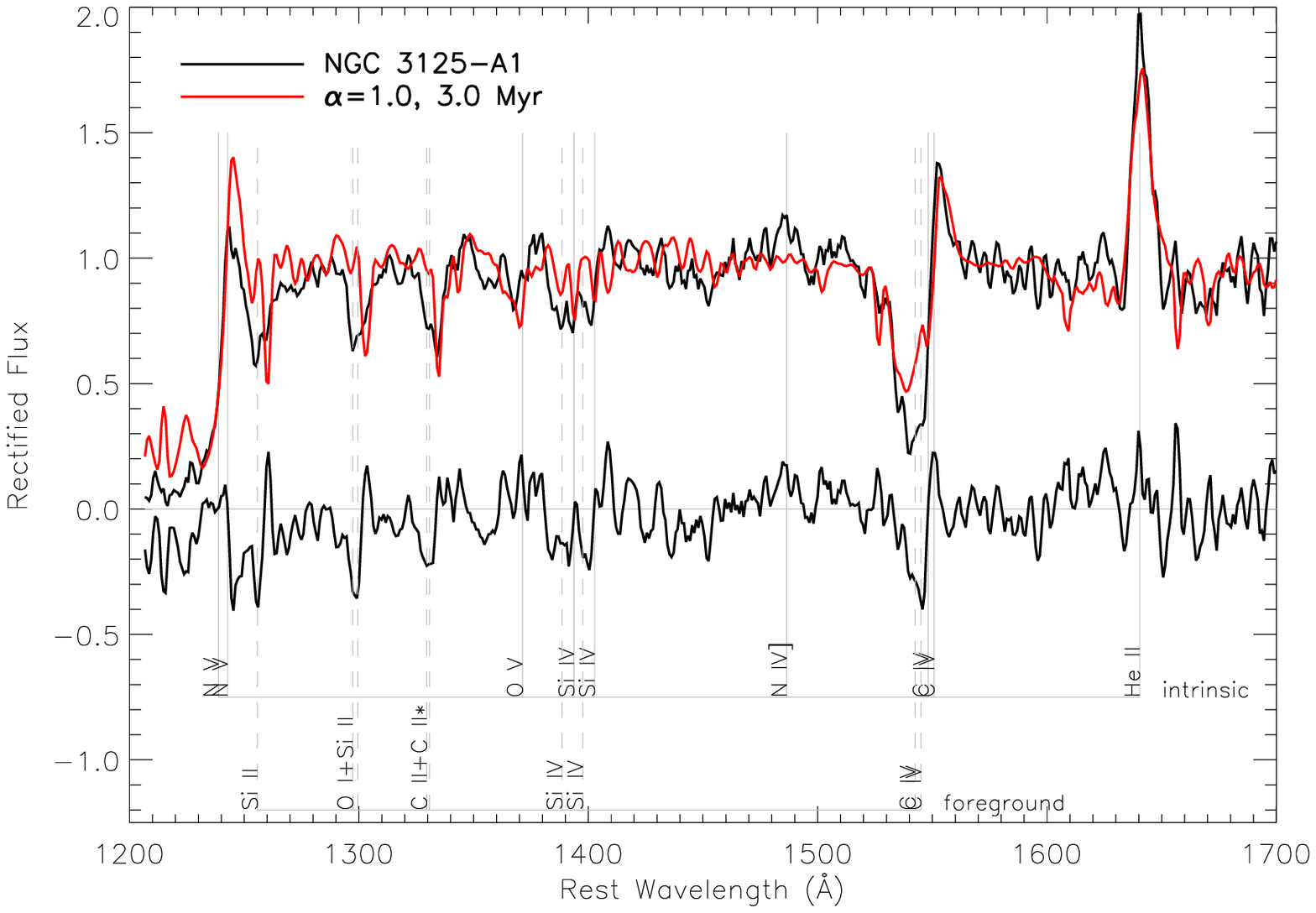}
\plotone{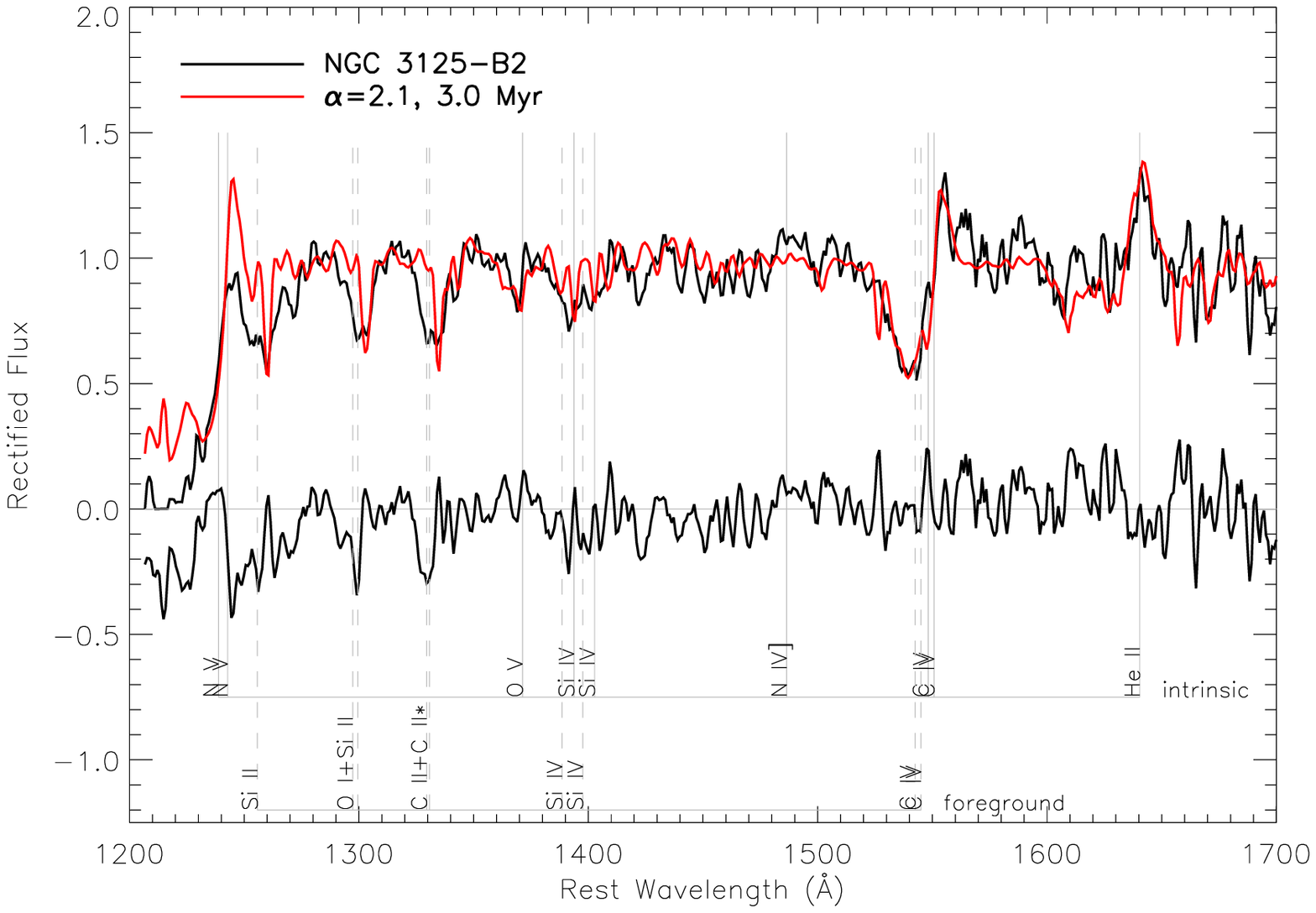}
\caption{Similar to Figure~\ref{fig7} but we fix the age at 3 Myr and use upper IMF exponents of $\alpha=$1.0/2.1 for A1/B2 respectively (left/right panels). The flatter than normal upper IMF exponents result in enhanced: i) O\vb~1371 absorption,  ii) He\ii~\lam1640 \AA~emission; and ii) N\vb~$\lambda$1238.8, 1242.8, Si\iv~$\lambda$1393.8, 1402.8, and C\iv~$\lambda$1548.2, 1550.8 P Cygni profiles. The residual curve in each panel shows that an extremely flat upper IMF exponent is required to fit the He\ii~1640 \AA~emission in A1, and that such an extreme exponent is not required for B2. Note that the observed N\iv] 1488 emission in A1 is still not reproduced by this model. Also note that the N\vb~1240 doublet is too strong in both cases. In summary, no single model can simultaneously reproduce features i), ii) and iii).}
\label{fig12}
\end{figure}

We performed a careful reddening correction of NGC 3125-A1 that brings the slope of the observed FUV-to-NUV spectroscopy in close agreement with the  expected slope from a model of appropriate age and metallicity. After this careful correction, we find that the strength of the  He\ii~$\lambda$1640 \AA~in NGC 3125-A1 is unprecedented on an absolute scale ($L=1.9\times10^{39}$ erg s$^{-1}$) and a relative scale ($EW=7.1\pm0.4$ \AA). There are two possible interpretations for the observed properties in A1, which we discuss next. 

i) The upper IMF exponent of A1 is flatter than normal. Adopting the Kroupa (2001) notation, this means that $\alpha<2.3$ for stars with masses between $0.5$ and 100 M$_\odot$, i.e., that the massive star content is enhanced. In order to show if this is a viable interpretation, we computed SSP models with upper IMF exponents of 1.0 for A1 and 2.1 for B2, keeping $\alpha=1.3$ for the lower stellar mass range (0.1-0.5 M$_\odot$). For the models, we select an age of 3 Myr, which corresponds to the time when the He\ii~1640 emission from WR stars is most prominent for a metallicity of $Z=0.020$. Recall that this is the metallicity for which we have predictions above 1600 \AA. In Figure~\ref{fig12} we compare the STIS UV observations of A1 and B2 with the above two models. The flatter upper IMF exponents produce stronger O\vb~1371 absorption and  provide a better match to the observed He\ii~1640 emission lines in both clusters. However, in order to match the strong He\ii~1640 emission in A1, the upper IMF exponent must be extremely flat. It would be even flatter at $Z=0.004$, the metallicity of NGC 3125. In addition, the flat exponent used for A1 produces a N\vb~1240 P Cygni profile that is excessively strong at the metallicity of NGC 3125, while the effects on Si\iv~1400 and C\iv~1550 are less noticeable. Finally, the  N\iv] 1488 emission from WR stars observed in A1 is not reproduced. Therefore, no single model can simultaneously reproduce all the observed features from massive stars in A1 and B2. In conclusion, we rule out a scenario where the upper IMF exponent of A1 is flat and equal to 1.0. Such exponent implies a fraction by mass of massive stars in A1 of $\sim$90\% when the typical observed value is $\sim$21\%. 

ii) In A1, the upper mass limit of the IMF is greater than the current limit of Starburst99 models, i.e., $>120$ M$_\odot$. Such massive stars have been found for instance in the 30 Doradus region of the LMC \citep{cro11}. Since very massive stars have short lifetimes ($<3$ Myr), they are rarely found. We could be witnessing a rare encounter with such stars in A1. At an age of 3-4 Myr, as previously derived, the very massive stars would have already died (Yusof et al. 2013). Starburst99 is not optimized for ages younger than 3 Myr, which would explain why for A1 we obtain older ages than expected if very massive stars are present.

The fact that O\vb~$\lambda$1371 is not usually detected in the spectra of nearby star forming galaxies but is present in A1 could be an independent indication of the presence of very massive stars. O\vb~\lam1371 absorption is present in the most massive stars of 30 Doradus, which also have strong He\ii~\lam1640 emission \citep{dek97}. Note that B2, which is not as extreme as A1, also shows O\vb~\lam1371 absorption, but in B2, the absorption is weaker.

In order to be able to reproduce the spectral signatures of peculiar populations such as that of 30 Doradus or NGC 3125-A1, stellar evolution tracks for very massive stars need to be implemented in the widely-used stellar population synthesis package, Starburst99. Such high-mass stellar evolution tracks are starting to become available and could be explored in the future.

Finally, Starburst99 assumes that massive stars evolve as single stars. However, it is newly appreciated that binarity is extremely common among massive stars \citep{mas09, san12, san13}.  Using their code, Binary Population and Spectral Synthesis (BPASS), \cite{eld12} show that the stellar He\ii~emission is boosted by the inclusion of the effect of massive stars being spun-up during binary mass transfer, as the latter results in these rapidly rotating stars experiencing quasi-homogeneous evolution. They also show that including this effect provides a better fit to the spectra of LBGs than assuming single star evolution. However, BPASS is still under development and the parameter space for binaries requires better constraints from observations, as pointed out by the authors. In addition, the equivalent width of the He\ii~1640 emission in NGC 3125-A1 is three times larger than what is observed in LBGs. Models including close binary and rotation cannot presently reproduce what is observed in NGC 3125-A1.

\section{SUMMARY \& CONCLUSION}

\begin{enumerate}
\item The UV spectrum of nearby blue compact dwarf galaxy NGC 3125 is relevant to the interpretation of higher redshift galaxy observations (C04). Thus, we aimed at characterizing the massive star distributions in its two main emitting regions. 
\item For this purpose, we obtained \hst~STIS spectra of compact star clusters NGC 3125-A1, B1, and B2; and a higher spectral resolution \hst~COS spectrum of the brightest UV cluster, A1. The STIS spectra span the wavelength ranges 1200-10,000 \AA~(A1 and B1) and 1200-3000 \AA~(B2), and have a nominal spectral resolution of $FWHM=0.9$ \AA~at 1300 \AA. The COS spectrum spans the wavelength range 1140-1450  \AA~and has a nominal spectral resolution of $FWHM=0.06$ \AA~at 1300 \AA.
\item A1 and B2 display broad He\ii~\lam1640 lines ($FWHM\gtrsim10^3$ km s$^{-1}$), originating in the winds of massive stars (Figure~ \ref{fig7}, Table~\ref{tab3}). A1 and B2 also show O\vb~$\lambda$1371 absorption. This line is rarely seen in the spectra of local star-forming galaxies and could be used as an independent indicator of the presence of a non-standard massive star content.
\item Using standard models, we derived ages of 3-4 Myr for the three clusters (Figure~\ref{fig7}).
\item The intrinsic reddenings of clusters A1, B1, and B2 are $E(B-V)_{\rm{g}}=0.13$, 0.15, and 0.13, respectively, as derived using the UV-continuum slope and SMC extinction law of \cite{gor03} (Tables~\ref{tab3} and~\ref{tab4}).
\item From the reddening-corrected luminosity of the UV stellar continuum, we obtained cluster masses of $1.7\times10^5$, $1.4\times10^5$, and $1.1\times10^5$ M$\odot$ for A1, B1, and B2, respectively.
\item The $N(WN5-6)/N(O)$ star ratios of clusters A1 and B2 are 0.23 and 0.10 respectively when using the above reddenings. These values are in agreement with previous measurements, but vary significantly when adopting reddenings derived from different methods (Table~\ref{tab5}).
\item The high $N(WN5-6)/N(O)$ star ratio obtained for A1, cannot be reproduced by current stellar population synthesis models that use a normal IMF and generic WR star line luminosities. We rule out a scenario where such ratio is due to a flat upper IMF. If this was the case, the N\vb~1240 P Cygni profile would be stronger (Figure~\ref{fig12}), and the IMF exponent of A1 would be unrealistically flat. At this moment, we cannot rule out another explanation, which is that in A1, the upper mass limit of the IMF is $>120$ M$_\odot$, i.e., greater than the current upper limit of the models. Such very massive hot stars have been found for instance in the 30 Doradus starburst of the LMC \citep{cro11}. The spectral similarity between A1 and very massive stars in the LMC suppornts this explanation. This would imply that A1 is younger than 3-4 Myr, which is the age inferred from models with an upper mass limit of $<120$ M$_\odot$.
\item Current models that include the effects of close binarity and rotation, which have been recently recognized as  important components of the evolution of massive stars, are currently unable to reproduce the strength of the He\ii~\lam1640 emission in A1, which is much stronger than is found for LBGs.
\item The take away point from our analysis is that in order to be able to reproduce peculiar stellar populations such as those of 30 Doradus and NGC 3125-A1, the present grid of stellar evolutionary tracks needs to be extended to higher masses than is currently available in stellar population synthesis packages such as Starburst99.
\end{enumerate}

\section{ACKNOWLEDGMENTS}

Support for program number 12172 was provided by NASA through a grant from the Space Telescope Science Institute, which is operated by the Association of Universities for Research in Astronomy, Inc., under NASA contract NAS5-26555. The research leading to these results has received funding from the European Research Council under the European Community's Seventh Framework Programme (FP7/2007-2013 Grant Agreement no. 321323). We thank the referee for the careful revision of this manuscript and for comments that greatly improved its quality.



\begin{deluxetable}{lllllllll}
\tablecolumns{9}
\tablewidth{0pc}
\tabletypesize{\scriptsize}
\tablecaption{Observations.}
\tablehead{Instr &  Type &  Aper &  Grating & $\lambda_{\rm{min}}$-$\lambda_{\rm{max}}$ & FWHM & Start Time & Stop Time & Exp Time \\ 
\hfill &  \hfill &   &  \hfill &  \AA &   \AA &  yr-mm-ds hh:mm:ss &  yr-mm-dd hh:mm:ss &  s \\
(1) &  (2) &  (3) &  (4) &  (5) &  (6) &  (7) &  (8) &  (9)}
\startdata                 
STIS &  Im &  28x50 &  MIRVIS & 5500-10,000 & - & 2011-11-20 10:43:40 & 2011-11-20 10:44:10 & 30.000 \\
STIS &  Sp &  25x2 &  G140L & 1150-1730  & 0.9 & 2012-01-26 21:48:54 & 2012-01-26 22:38:21 & 2967.000 \\
STIS &  Sp &  25x2 &  G230L & 1570-3180  & 2.4 & 2012-01-26 20:49:05 & 2012-01-26 21:02:33 & 808.000 \\
STIS &  Sp &  52x2 &  G430L & 2900-5700  & 4.1 & 2012-01-26 20:37:17 & 2012-01-26 20:42:45 & 240.000 \\
STIS &  Sp &  52x2 &  G750L & 5240-10,270  & 7.4 & 2012-01-26 20:23:03 & 2012-01-26 20:26:31 & 120.000 \\
COS &  Im &  2.5 &  MIRRORA & 1650-3200 & -  & 2012-01-26 20:14:15 & 2012-01-26 20:18:30 & 40.1 \\
COS &  Sp &  2.5 &  G130M & 1144-1441  & 0.06 & 2011-11-20 10:49:22 & 2011-11-20 11:31:51 & 2549.184 \\
COS & Sp & 2.5 & G130M & 1163-1460 & 0.06 & 2011-11-20 12:14:39 & 2011-11-20 13:04:10 & 2971.200 \\
\enddata\\[-15pt]	
\tablecomments{(1) Instrument.  (2) Observation type. Im=Image. Sp=Spectrum.  (3) Aperture FOV (after projection onto the detector, if applicable).  (4) Grating (if applicable).  (5) Wavelength coverage.  (6) Nominal spectral resolution.  (7)/(8)/(9) Start/stop/exposure time of observation.}		
\label{tab1}
\end{deluxetable}


\clearpage
\begin{deluxetable}{llllll}
\tablecolumns{6}
\tablewidth{0pc}
\tablecaption{Presence of WR Star Lines.}
\tablehead{Subtype	&	Feature	&	Other designation	&	A1	&	B1	&	B2}
\startdata 
Subtype	&	Feature	&	Other designation	&	A1	&	B1	&	B2	\\
WN	&	N\iv~\lam1488	&	\nodata	&	1	&	0	&	0	\\
WN+WC	&	He\ii~\lam1640	&	\nodata	&	1	&	1	&	1	\\
WN	&	N\vb~\lam1718	&	\nodata	&	N/A	&	N/A	&	N/A	\\
WN+WC	&	N\iii~\lam\lam4634, 4640	&	N\iii~\lam4640/C\iii~\lam4650 bump	&	0	&	0	&	N/A	\\
WN	&	He\ii~\lam4686	&	\nodata	&	1	&	0	&	N/A	\\
WC	&	C\iv~\lam5801, 5813	&	C\iv~\lam5808	&	0	&	0	&	N/A	\\
\enddata\\[-15pt]
\label{tab2}
\end{deluxetable}


\begin{deluxetable}{lllll}
\tablecolumns{5}
\tablewidth{0pc}
\tabletypesize{\scriptsize}
\tablecaption{Reddening.}
\tablehead{\hfill	&	A1	&	B1	&	B2	&	Units}
\startdata
1	$F_{\rm{1640}}$	&	1.8E-14	&	4.2E-15	&	4.5E-15	&	erg s$^{-1}$ cm$^{-2}$	\\
2	$F_{\rm{4686}}$	&	4.9E-15	&	\nodata	&	\nodata	&	erg s$^{-1}$ cm$^{-2}$	\\
3	$F_{\rm{4861}}$	&	4.3E-14	&	\nodata	&	\nodata	&	erg s$^{-1}$ cm$^{-2}$	\\
4	$F_{\rm{6563}}$	&	1.8E-13	&	\nodata	&	\nodata	&	erg s$^{-1}$ cm$^{-2}$	\\
5	$EW_{1640}$	&	7.1	&	\nodata	&	2.5	&	\AA	\\
6	$EW_{4686}$	&	14.2	&	\nodata	&	\nodata	&	\AA	\\
7	$E(B-V)_{\rm{G}}$	&	0.08	&	0.08	&	0.08	&	mag	\\
8	$E(B-V)_{\rm{g, \beta, smc}}$	&	0.13	&	0.15	&	0.13	&	mag	\\
9	$E(B-V)_{\rm{g, \beta, sbt}}$	&	0.34	&	0.40	&	0.33	&	mag	\\
10	$E(B-V)_{\rm{g, h, smc}}$	&	0.26	&	\nodata	&	\nodata	&	mag	\\
11	$E(B-V)_{\rm{g, h, sbt}}$	&	0.24	&	\nodata	&	\nodata	&	mag	\\
12	$E(B-V)_{\rm{g, he, smc}}$	&	0.10	&	\nodata	&	\nodata	&	mag	\\
13	$E(B-V)_{\rm{g, he, sbt}}$	&	0.16	&	\nodata	&	\nodata	&	mag	\\
14	$\beta_{\rm{g,\beta,smc}}$	&	-2.56	&	-2.60	&	-2.64	&	\nodata	\\
15	$\beta_{\rm{g,\beta,sbt}}$	&	-2.63	&	-2.64	&	-2.59	&	\nodata	\\
16	$(L_{\rm{1640}}/L_{\rm{4686}})_{\rm{he,smc}}$	&	10.0	&	\nodata	&	\nodata	&	\nodata	\\
17	$(L_{\rm{1640}}/L_{\rm{4686}})_{\rm{he,sbt}}$	&	10.1	&	\nodata	&	\nodata	&	\nodata	\\
18	$(L_{\rm{6563}}/L_{\rm{4861}})_{\rm{h,smc}}$	&	2.85	&	\nodata	&	\nodata	&	\nodata	\\
19	$(L_{\rm{6563}}/L_{\rm{4861}})_{\rm{h,sbt}}$	&	2.85	&	\nodata	&	\nodata	&	\nodata	\\
\enddata\\[-5pt]	
\tablecomments{$\beta$=We de-reddened using the UV-slope method. h=We de-reddened using the Balmer-line ratio method. he=We de-reddened using the He\ii-line ratio method. smc=We used the SMC extinction curve of \cite{gor03}. sbt=We used the starburst attenuation curve of \cite{cal94}. 1-4. Reddening-uncorrected line fluxes. 5-6. Line equivalent widths corrected for the instrumental profile. 7. Galactic reddening. 8-13. Intrinsic reddening from the UV-slope, $E(B-V)_{\rm{g, \beta}}$, the Balmer-line ratio, $E(B-V)_{\rm{g, h}}$, and the He\ii-line ratio , $E(B-V)_{\rm{g, he}}$. 14-15 $\beta$-slope obtained after applying a reddening correction of $E(B-V)_{\rm{G}}$+$E(B-V)_{\rm{g, \beta}}$. 16-17. He\ii~line ratio obtained after applying a reddening correction of $E(B-V)_{\rm{G}}$+$E(B-V)_{\rm{g,he}}$. 18-19. Balmer-line ratio obtained after applying a reddening correction of $E(B-V)_{\rm{G}}$+$E(B-V)_{\rm{g,h}}$.}
\label{tab3}
\end{deluxetable}


\begin{deluxetable}{lllll}
\tablecolumns{5}
\tablewidth{0pc}
\tabletypesize{\scriptsize}
\tablecaption{Luminosities and Masses.}
\tablehead{\hfill	&	A1	&	B1	&	B2	&	Units}
\startdata
1	$L_{\rm{1500, \beta, smc}}$	&	2.6E+38	&	2.2E+38	&	1.7E+38	&	erg s$^{-1}$	\\
2	$L_{\rm{1500, \beta, sbt}}$	&	1.4E+39	&	1.6E+39	&	8.3E+38	&	erg s$^{-1}$	\\
3	$L_{\rm{1500, h, smc}}$	&	1.2E+39	&	\nodata	&	\nodata	&	erg s$^{-1}$	\\
4	$L_{\rm{1500, h, sbt}}$	&	5.3E+38	&	\nodata	&	\nodata	&	erg s$^{-1}$	\\
5	$L_{\rm{1500, he, smc}}$	&	1.8E+38	&	\nodata	&	\nodata	&	erg s$^{-1}$	\\
6	$L_{\rm{1500, he, sbt}}$	&	2.5E+38	&	\nodata	&	\nodata	&	erg s$^{-1}$	\\
7	$L_{\rm{1640, \beta, smc}}$	&	1.9E+39	&	\nodata	&	5.0E+38	&	erg s$^{-1}$	\\
8	$L_{\rm{1640, \beta, sbt}}$	&	1.1E+40	&	\nodata	&	3.9E+39	&	erg s$^{-1}$	\\
9	$L_{\rm{1640,h, smc}}$	&	7.7E+39	&	\nodata	&	\nodata	&	erg s$^{-1}$	\\
10	$L_{\rm{1640,h, sbt}}$	&	4.3E+39	&	\nodata	&	\nodata	&	erg s$^{-1}$	\\
11	$L_{\rm{1640,he, smc}}$	&	1.4E+39	&	\nodata	&	\nodata	&	erg s$^{-1}$	\\
12	$L_{\rm{1640,he, sbt}}$	&	2.1E+39	&	\nodata	&	\nodata	&	erg s$^{-1}$	\\
13	$L_{\rm{4686,\beta, smc}}$	&	1.6E+38	&	\nodata	&	\nodata	&	erg s$^{-1}$	\\
14	$L_{\rm{4686,\beta, sbt}}$	&	4.5E+38	&	\nodata	&	\nodata	&	erg s$^{-1}$	\\
15	$L_{\rm{4686, h, smc}}$	&	3.5E+38	&	\nodata	&	\nodata	&	erg s$^{-1}$	\\
16	$L_{\rm{4686, h, sbt}}$	&	4.4E+38	&	\nodata	&	\nodata	&	erg s$^{-1}$	\\
17	$L_{\rm{4686, he, smc}}$	&	1.4E+38	&	\nodata	&	\nodata	&	erg s$^{-1}$	\\
18	$L_{\rm{4686, he, sbt}}$	&	2.1E+38	&	\nodata	&	\nodata	&	erg s$^{-1}$	\\
19	$L_{\rm{4861,\beta,smc}}$	&	1.3E+39	&	\nodata	&	\nodata	&	erg s$^{-1}$	\\
20	$L_{\rm{4861,\beta,sbt}}$	&	3.7E+39	&	\nodata	&	\nodata	&	erg s$^{-1}$	\\
21	$L_{\rm{4861,h,smc}}$	&	2.0E+39	&	\nodata	&	\nodata	&	erg s$^{-1}$	\\
22	$L_{\rm{4861,h,sbt}}$	&	2.4E+39	&	\nodata	&	\nodata	&	erg s$^{-1}$	\\
23	$L_{\rm{4861,he,smc}}$	&	1.2E+39	&	\nodata	&	\nodata	&	erg s$^{-1}$	\\
24	$L_{\rm{4861,he,sbt}}$	&	1.7E+39	&	\nodata	&	\nodata	&	erg s$^{-1}$	\\
25	$L_{\rm{6563,\beta,smc}}$	&	4.3E+39	&	\nodata	&	\nodata	&	erg s$^{-1}$	\\
26	$L_{\rm{6563,\beta,sbt}}$	&	9.4E+39	&	\nodata	&	\nodata	&	erg s$^{-1}$	\\
27	$L_{\rm{6563,h,smc}}$	&	5.6E+39	&	\nodata	&	\nodata	&	erg s$^{-1}$	\\
28	$L_{\rm{6563,h,sbt}}$	&	7.0E+39	&	\nodata	&	\nodata	&	erg s$^{-1}$	\\
29	$L_{\rm{6563,he,smc}}$	&	4.1E+39	&	\nodata	&	\nodata	&	erg s$^{-1}$	\\
30	$L_{\rm{6563,he,sbt}}$	&	5.4E+39	&	\nodata	&	\nodata	&	erg s$^{-1}$	\\
31	$M_{\rm{*,1500,\beta,smc}}$	&	1.7E+05	&	1.4E+05	&	1.1E+05	&	M$\odot$	\\
32	$M_{\rm{*,1500,\beta,sbt}}$	&	8.9E+05	&	1.0E+06	&	5.4E+05	&	M$\odot$	\\
33	$M_{\rm{*,1500,h,smc}}$	&	8.1E+05	&	\nodata	&	\nodata	&	M$\odot$	\\
34	$M_{\rm{*,1500, h, sbt}}$	&	3.4E+05	&	\nodata	&	\nodata	&	M$\odot$	\\
35	$M_{\rm{*,1500, he, smc}}$	&	1.2E+05	&	\nodata	&	\nodata	&	M$\odot$	\\
36	$M_{\rm{*,1500, he, sbt}}$	&	1.6E+05	&	\nodata	&	\nodata	&	M$\odot$	\\
\enddata\\[-5pt]	
\tablecomments{1-30. Reddening-corrected line luminosities. 31-36. Masses derived when different reddening-corrections are applied. The de-reddening method and reddening curve labels are as in Table~\ref{tab3}.}
\label{tab4}
\end{deluxetable}


\begin{deluxetable}{lllll}
\tablecolumns{5}
\tablewidth{0pc}
\tabletypesize{\tiny}
\tablecaption{$N(WR)/N(O)$ ratios.}
\tablehead{\hfill	&	A1	&	B1	&	B2	&	Units}
\startdata
1	$N(O)_{\rm{1500,\beta,smc}}$	&	467	&	351	&	280	&	stars	\\
2	$N(O)_{\rm{1500,\beta,sbt}}$	&	2494	&	2518	&	1326	&	stars	\\
3	$N(O)_{\rm{1500,h,smc}}$	&	2264	&	\nodata	&	\nodata	&	stars	\\
4	$N(O)_{\rm{1500,h,sbt}}$	&	960	&	\nodata	&	\nodata	&	stars	\\
5	$N(O)_{\rm{1500,he,smc}}$	&	326	&	\nodata	&	\nodata	&	stars	\\
6	$N(O)_{\rm{1500,he,sbt}}$	&	447	&	\nodata	&	\nodata	&	stars	\\
7	$N(O)_{\rm{H\beta,\beta,smc}}$	&	236	&	\nodata	&	\nodata	&	stars	\\
8	$N(O)_{\rm{H\beta,\beta,sbt}}$	&	668	&	\nodata	&	\nodata	&	stars	\\
9	$N(O)_{\rm{H\beta,h,smc}}$	&	352	&	\nodata	&	\nodata	&	stars	\\
10	$N(O)_{\rm{H\beta,h,sbt}}$	&	438	&	\nodata	&	\nodata	&	stars	\\
11	$N(O)_{\rm{H\beta,he,smc}}$	&	215	&	\nodata	&	\nodata	&	stars	\\
12	$N(O)_{\rm{H\beta,he,sbt}}$	&	312	&	\nodata	&	\nodata	&	stars	\\
13	$N(WN5-6)_{\rm{1640,\beta,smc}}$	&	108	&	\nodata	&	28	&	stars	\\
14	$N(WN5-6)_{\rm{1640,\beta,sbt}}$	&	589	&	\nodata	&	218	&	stars	\\
15	$N(WN5-6)_{\rm{1640,h,smc}}$	&	428	&	\nodata	&	\nodata	&	stars	\\
16	$N(WN5-6)_{\rm{1640,h,sbt}}$	&	238	&	\nodata	&	\nodata	&	stars	\\
17	$N(WN5-6)_{\rm{1640,he,smc}}$	&	78	&	\nodata	&	\nodata	&	stars	\\
18	$N(WN5-6)_{\rm{1640,he,sbt}}$	&	115	&	\nodata	&	\nodata	&	stars	\\
19	$N(WN5-6)_{\rm{4686,\beta,smc}}$	&	86	&	\nodata	&	\nodata	&	stars	\\
20	$N(WN5-6)_{\rm{4686,\beta,sbt}}$	&	253	&	\nodata	&	\nodata	&	stars	\\
21	$N(WN5-6)_{\rm{4686,h,smc}}$	&	197	&	\nodata	&	\nodata	&	stars	\\
22	$N(WN5-6)_{\rm{4686,h,sbt}}$	&	244	&	\nodata	&	\nodata	&	stars	\\
23	$N(WN5-6)_{\rm{4686,he,smc}}$	&	78	&	\nodata	&	\nodata	&	stars	\\
24	$N(WN5-6)_{\rm{4686,he,sbt}}$	&	115	&	\nodata	&	\nodata	&	stars	\\
25	$N(WN5-6)/N(O)_{\rm{1640,1500,\beta,\beta, smc}}$	&	0.23	&	\nodata	&	0.10	&	\nodata	\\
26	$N(WN5-6)/N(O)_{\rm{1640,1500,\beta,\beta, sbt}}$	&	0.24	&	\nodata	&	0.16	&	\nodata	\\
27	$N(WN5-6)/N(O)_{\rm{1640,1500,h,h, smc}}$	&	0.19	&	\nodata	&	\nodata	&	\nodata	\\
28	$N(WN5-6)/N(O)_{\rm{1640,1500,h,h, sbt}}$	&	0.25	&	\nodata	&	\nodata	&	\nodata	\\
29	$N(WN5-6)/N(O)_{\rm{1640,1500,he,he, smc}}$	&	0.24	&	\nodata	&	\nodata	&	\nodata	\\
30	$N(WN5-6)/N(O)_{\rm{1640,1500,he,he, sbt}}$	&	0.26	&	\nodata	&	\nodata	&	\nodata	\\
31	$N(WN5-6)/N(O)_{\rm{4686,H\beta,\beta,\beta, smc}}$	&	0.37	&	\nodata	&	\nodata	&	\nodata	\\
32	$N(WN5-6)/N(O)_{\rm{4686,H\beta,\beta,\beta, sbt}}$	&	0.38	&	\nodata	&	\nodata	&	\nodata	\\
33	$N(WN5-6)/N(O)_{\rm{4686,H\beta,h,h, smc}}$	&	0.56	&	\nodata	&	\nodata	&	\nodata	\\
34	$N(WN5-6)/N(O)_{\rm{4686,H\beta,h,h, sbt}}$	&	0.56	&	\nodata	&	\nodata	&	\nodata	\\
35	$N(WN5-6)/N(O)_{\rm{4686,H\beta,he,he, smc}}$	&	0.36	&	\nodata	&	\nodata	&	\nodata	\\
36	$N(WN5-6)/N(O)_{\rm{4686,H\beta,he,he, sbt}}$	&	0.37	&	\nodata	&	\nodata	&	\nodata	\\
37	$N(WN5-6)/N(O)_{\rm{min}}$	&	0.19	&	\nodata	&	0.10	&	\nodata	\\
38	$N(WN5-6)/N(O)_{\rm{max}}$	&	0.56	&	\nodata	&	0.16	&	\nodata	\\
39	$N(WN5-6)/N(O)_{\rm{1640,1500,he/\beta, smc}}$	&	0.17	&	\nodata	&	\nodata	&	\nodata	\\
40	$N(WN5-6)/N(O)_{\rm{1640,1500,he,\beta, sbt}}$	&	0.05	&	\nodata	&	\nodata	&	\nodata	\\
41	$N(WN5-6)/N(O)_{\rm{4686,H\beta,he,h, smc}}$	&	0.22	&	\nodata	&	\nodata	&	\nodata	\\
42	$N(WN5-6)/N(O)_{\rm{4686,H\beta,he,h, sbt}}$	&	0.26	&	\nodata	&	\nodata	&	\nodata	\\
\enddata\\[-5pt]	
\tablecomments{1-12.  $N(O)$ derived by scaling the expected $N(O)$ by either $L_{1500}$ or $L_{4861}$. We use $N(O)=2793$ and $L_{4861}=10^{40.193}$ erg s$^{-1}$, which correspond to an SSP of 3 Myr, $M_{\rm{*}}=10^6$ M$\odot$, $Z=0.004$ and \cite{kro01} IMF. At 4 Myr, these values are$N(O)= 2470$ and $L_{4861}=10^{39.972}$ erg s$^{-1}$. 13-24. $N(WN5-6)$ derived by dividing the reddening corrected 1640 and 4686~\AA He\ii-line luminosities by template luminosities of $1.8\times10^{37}$ erg s$^{-1}$ and $1.8\times10^{36}$  erg s$^{-1}$, respectively from \cite{cro06}. 25-36. $N(WN5-6)/N(O)$ ratios. 37-38. Minimum and maximum ratios. 39-42. $N(WN5-6)/N(O)$ ratios obtained when different reddenings are used for $N(O)$ and $N(WN5-6)$. The de-reddening method and reddening curve labels are as in Table~\ref{tab3}.}
\label{tab5}
\end{deluxetable}


\clearpage

\end{document}